\documentclass[twocolumn,showpacs,prb,superscriptaddress,a4paper,floatfix]{revtex4}
\usepackage{amssymb}
\usepackage{graphicx}
\usepackage{amsmath}
\usepackage{graphicx}
\usepackage{dcolumn}
\usepackage{bm}

\begin{document}

\title{Modeling electronic structure and transport properties of graphene
with resonant scattering centers}
\author{Shengjun Yuan}
\email{s.yuan@science.ru.nl}
\affiliation{Institute for Molecules and Materials, Radboud University of Nijmegen,
NL-6525ED Nijmegen, The Netherlands}
\author{Hans De Raedt}
\email{h.a.de.raedt@rug.nl}
\affiliation{Department of Applied Physics, Zernike Institute for Advanced Materials,
University of Groningen, Nijenborgh 4, NL-9747AG Groningen, The Netherlands}
\author{Mikhail I. Katsnelson}
\email{m.katsnelson@science.ru.nl}
\affiliation{Institute for Molecules and Materials, Radboud University of Nijmegen,
NL-6525ED Nijmegen, The Netherlands}
\date{\today }

\begin{abstract}
We present a detailed numerical study of the electronic properties of
single-layer graphene with resonant (\textquotedblleft
hydrogen\textquotedblright ) impurities and vacancies within a framework of
noninteracting tight-binding model on a honeycomb lattice. The algorithms
are based on the numerical solution of the time-dependent Schr\"{o}dinger
equation and applied to calculate the density of states, \textit{\
quasieigenstates}, AC and DC conductivities of large samples containing
millions of atoms. Our results give a consistent picture of evolution of
electronic structure and transport properties of functionalized graphene in
a broad range of concentration of impurities (from graphene to graphane),
and show that the formation of impurity band is the main factor determining
electrical and optical properties at intermediate impurity concentrations,
together with a gap opening when approaching the graphane limit.
\end{abstract}

\pacs{72.80.Vp, 73.22.Pr, 78.67.Wj}
\maketitle

\section{Introduction}

The experimental realization of a single layer of carbon atoms arranged in a
honeycomb lattice (graphene) has prompted huge activity in both experimental
and theoretical physics communities (for reviews, see Refs. %
\onlinecite{r1,r2,r3,r4,r5,r6,r7,Cresti2008,Mucciolo2010,Peres2010}).
Graphene in real experiments always has different kinds of disorder or
impurities, such as ripples, adatoms, admolecules, etc. One of the most
important problems in graphene physics, especially, keeping in mind
potential applications of graphene in electronics, is understanding the
effect of these imperfections on the electronic structure and transport
properties.

Being massless Dirac fermions with the wavelength much larger than the
interatomic distance, charge carriers in graphene scatter rather weakly by
generic short-range scattering centers, similar to weak light scattering
from obstacles with sizes much smaller that the wavelength. The scattering
theory for Dirac electrons in two dimensions is discussed in Refs. %
\onlinecite{Shon1998,Katsnelson2007,Hentschel2007,Novikov2007}. Long-range
scattering centers are of special importance for transport properties, such
as charge impurities \cite{r6,Nomura2006,Ando2006,Hwang2007}, ripples
created long-range elastic deformations \cite{r7,Katsnelson2008}, and
resonant scattering centers \cite%
{Peres2006,Katsnelson2007,Katsnelson2008,Ostrovsky2006,Stauber2007,Titov2010}%
. In the latter case, the divergence of the scattering length provides a
long-range scattering and a very slow, logarithmic, decay of the scattering
phase near the Dirac (neutrality) point. Earlier the resonant scattering of
Dirac fermions was studied in a context of $d$-wave high-temperature
superconductivity \cite{Altland2002}. For the case of graphene, vacancies
are prototype examples of the resonant scatterers \cite{Stauber2007,Chen2009}%
. Numerous adatoms and admolecules (including the important case of hydrogen
atoms covalently bonded with carbon atoms) provide other examples \cite%
{Wehling2008,Wehling2009,Wehling2009b}. Recently, some experimental \cite%
{Ni2010} and theoretical \cite{Wehling2010} evidence appeared that,
probably, the resonant scattering due to carbon-carbon bonds between organic
admolecules and graphene is the main restricting factor for electron
mobility in graphene on a substrate. Resonant scattering also plays an
important role in interatomic interactions and ordering of adatoms on
graphene \cite{Shytov2009}. This all makes the theoretical study of graphene
with resonant scattering centers an important problem.

In the present paper, we study this issue by direct numerical simulations of
electrons on a honeycomb lattice in the framework of the tight-binding
model. Numerical calculations based on exact diagonalization can only treat
samples with relative small number of sites, for example, to study the
quasilocalization of eigenstate close to the neutrality point around the
vacancy \cite{Peres2006,Pereira2008} and the splitting of zero-energy Landau
levels in the presence of random nearest neighbor hoping \cite{Pereira2009}.
For large graphene sheet with millions of atoms, the numerical calculation
of an important property, the density of states (DOS), is mainly performed
by the recursion method \cite{Pereira2008,Pereira2008b,Pereira2006} and
time-evolution method \cite{DeRaedt2008,Wehling2010}. The time-evolution
method is based on numerical solution of time-dependent Schr\"{o}dinger
equation with additional averaging over random superposition of basis
states. In this paper, we extend the method of Ref. \onlinecite{Hams2000} to
compute the eigenvalue distribution of very large matrices to the
calculation of transport coefficients. It allows us to carry out
calculations for rather large systems, up to hundreds of millions of sites,
with a computational effort that increases only linearly with the system
size. Furthermore, another extension of the time-evolution method yields the 
\textit{quasieigenstate}, a random superposition of degenerate energy
eigenstates, as well as the AC and DC~\cite{Wehling2010} conductivities.

The numerical calculation of the conductivity is based on the Kubo formula
of noninteracting electrons. The details of these algorithm will be given in
this paper. Our numerical results are consistent with the results on
hydrogenated graphene~\cite{Bang2010} and graphene with vacancies~\cite%
{Wu2010}, which are based on the numerical calculation of the Kubo-Greenwood
formula~\cite{Roche1997}. Another widely used method of the numerical study
of electronic transport in graphene is the recursive Green's function method 
\cite%
{Cresti2007,Lewenkopf2007,Zhu2010,Hilke2009,ZhangYY2008,LongW2008,ZhangYY2009,ZhangYY2009b,Lherbier2008,Mucciolo2009}%
, which is generally applied to relatively small samples followed by
averaging of many different configurations. The recursive Green's function
method is a powerful tool to calculate the electronic transport in small
system such as graphene ribbons, while the method that we employ in this
paper is more suitable for large systems having millions of atoms and
therefore does not involve averaging over different realizations.

The paper is organized as follows. Section II gives a description of the
tight-binding Hamiltonian of single layer graphene including different types
of disorders or impurities, in the absence and presence of a perpendicular
magnetic field. In section III, we first discuss briefly the numerical
method used to calculate the DOS, and show the accuracy of this algorithm by
comparing the analytical and numerical results for clean graphene. Then,
based on the calculation the DOS, we discuss the effects of vacancies or
resonant impurities to the electronic structure of graphene, including the
broadening of the Landau levels and the split of zero Landau levels. In
section IV, we introduce the concept of a \textit{quasieigenstates}, and use
it to show the quasilocalization of the states around the vacancies or
resonant impurities. Sections V and VI give discussions of the AC and DC
conductivities, respectively. The details of numerical methods and various
examples are discussed in detail in each section. Finally a brief general
discussion is given in section VII.

\section{Tight-binding model}

The tight-binding Hamiltonian of a single-layer graphene is given by%
\begin{equation}
H=H_{0}+H_{1}+H_{v}+H_{imp},  \label{Hamiltonian}
\end{equation}%
where $H_{0}$ derives from the nearest neighbor interactions of the carbon
atoms:%
\begin{equation}
H_{0}=-\sum_{<i,j>}t_{ij}c_{i}^{+}c_{j},
\end{equation}%
$H_{1}$ represents the next-nearest neighbor interactions of the carbon
atoms:%
\begin{equation}
H_{1}=-\sum_{<<i,j>>}t_{ij}^{\prime }c_{i}^{+}c_{j},
\end{equation}%
$H_{v}$ denotes the on-site potential of the carbon atoms:%
\begin{equation}
H_{v}=\sum_{i}v_{i}c_{i}^{+}c_{i},
\end{equation}%
and $H_{imp}$ describes the resonant impurities:%
\begin{equation}
H_{imp}=\varepsilon _{d}\sum_{i}d_{i}^{+}d_{i}+V\sum_{i}\left(
d_{i}^{+}c_{i}+H.c.\right) .
\end{equation}%
For discussions of the last term see, e.g. Refs.~%
\onlinecite{Wehling2009b,Robinson2008}.

\begin{figure}[t]
\begin{center}
\includegraphics[width=8cm]{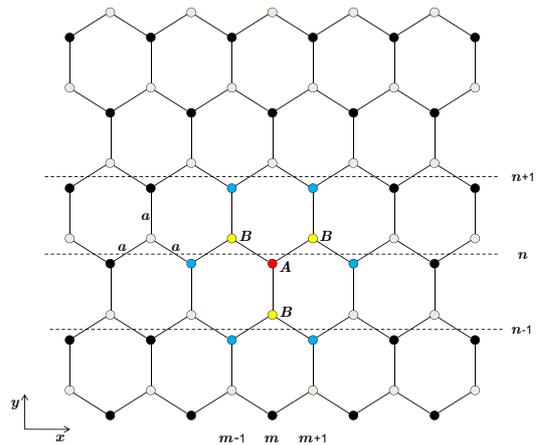}
\end{center}
\caption{(Color online) The lattice structure of a graphene sheet. Each
carbon is labeled by an coordinate $\left( m,n\right) $, where $m$ is along
the zigzag edge and $n$ is along the armchair edge. Each carbon (red) has
three nearest neighbors (yellow) and six next-nearest neighbors (blue).}
\label{graphenepic}
\end{figure}
The spin degree of freedom contributes only through a degeneracy factor and
is omitted for simplicity in Eq.~(\ref{Hamiltonian}). Vacancies are
introduced by simply removing the corresponding carbon atoms from the sample.

If a magnetic field is applied to the graphene layer, the hopping integrals
are replaced by a Peierls substitution \cite{Vonsovsky1989}, that is, the
hopping parameter becomes 
\begin{equation}
t_{mn}\rightarrow t_{mn}e^{ie\int_{m}^{n}\mathbf{A}\cdot d\mathbf{l}%
}=t_{mn}e^{i\left( 2\pi /\Phi _{0}\right) \int_{m}^{n}\mathbf{A}\cdot d%
\mathbf{l}},
\end{equation}%
where $\int_{m}^{n}\mathbf{A}\cdot d\mathbf{l}$ is the line integral of the
vector potential from site $m$ to site $n$, and the flux quantum $\Phi
_{0}=ch/e$.

Consider a single graphene layer with a perpendicular magnetic field $%
\mathbf{B}=(0,0,B)$. Let the zigzag edge be along the $x$ axis, and use the
Landau gauge, that is, the vector potential $\mathbf{A}=(-By,0,0)$ Then $%
H_{0}$\ changes into 
\begin{eqnarray}
H_{0} &=&\sum_{m,n}t_{\left( m,n\right) ,\left( m,n-1\right)
}a_{m,n}^{+}b_{m,n-1}  \notag \\
&&+t_{\left( m,n\right) ,\left( m-1,n\right) }e^{i\pi n(\Phi /\Phi
_{0})}a_{m,n}^{+}b_{m-1,n}  \notag \\
&&+t_{\left( m,n\right) ,\left( m+1,n\right) }e^{-i\pi n(\Phi /\Phi
_{0})}a_{m,n}^{+}b_{m+1,n}  \notag \\
&&+H.c.
\end{eqnarray}%
where%
\begin{equation}
\Phi \equiv \frac{3\sqrt{3}}{2}Ba^{2},
\end{equation}
$a$ is the nearest-neighbor interatomic distance.

\section{Density of States}

\begin{figure}[t]
\begin{center}
\includegraphics[width=8cm]{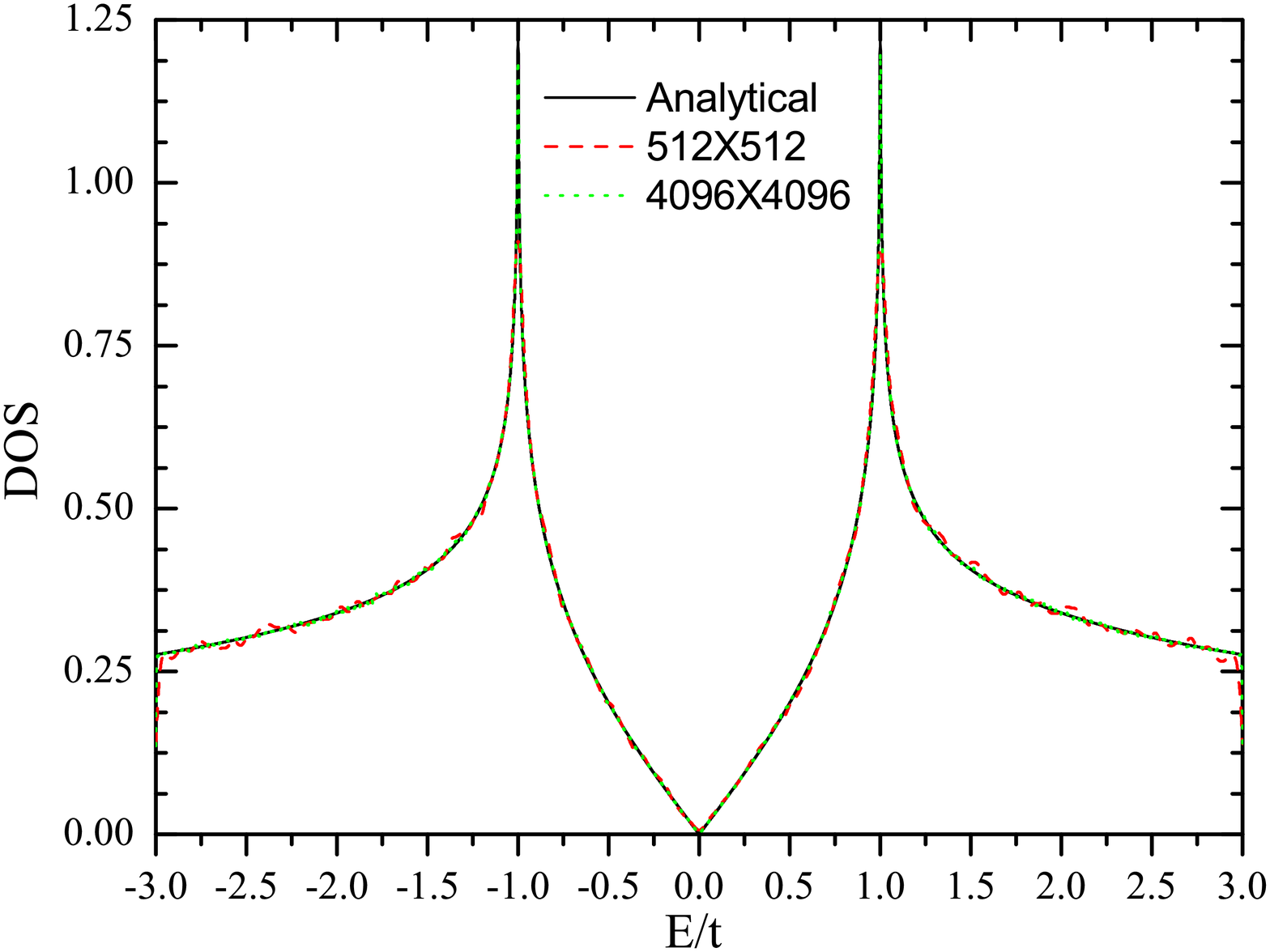}
\end{center}
\caption{(Color online) Comparison of the analytical DOS (in units of $1/t$, black solid) with
the numerical results of a sample contains $512\times 512$ (red dash) or $%
4096\times 4096$ (green dot) carbon atoms.}
\label{doscomparesize}
\end{figure}

\begin{figure*}[t]
\begin{center}
\includegraphics[width=16cm]{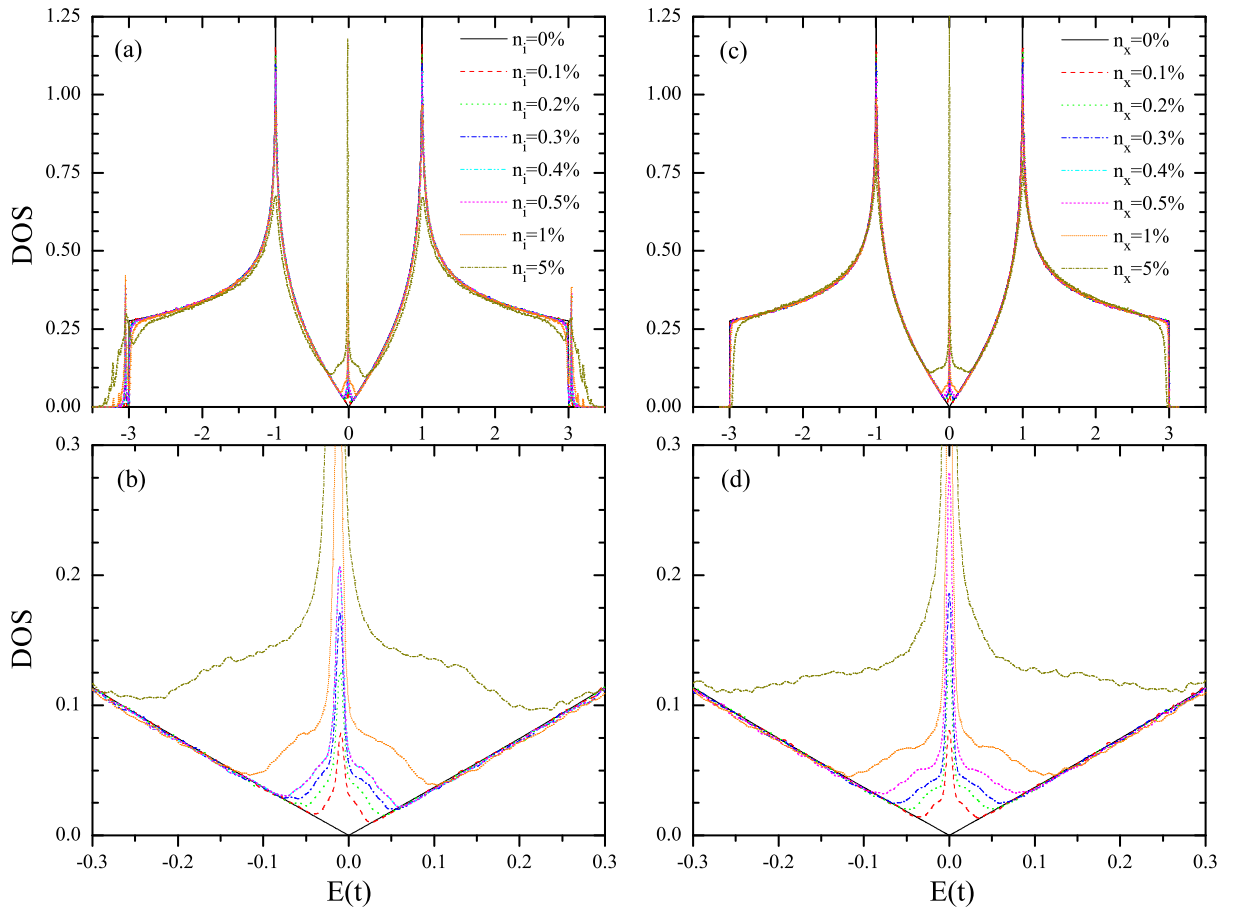}
\end{center}
\caption{(Color online) Density of states (in units of $1/t$) as a function of energy $E$ (in
units of $t$) for different resonant impurity ($\protect\varepsilon %
_{d}=-t/16,$ $V=2t$) or vacancy concentrations: $n_{i}(n_{x})=0.1\%,$ $%
0.2\%, $ $0.3\%,$ $0.4\%,$ $0.5\%,$ $1\%,$ $5\%$. Sample size is $4096\times
4096.$}
\label{dosximpandx}
\end{figure*}

\begin{figure}[t]
\begin{center}
\mbox{
\includegraphics[width=8cm]{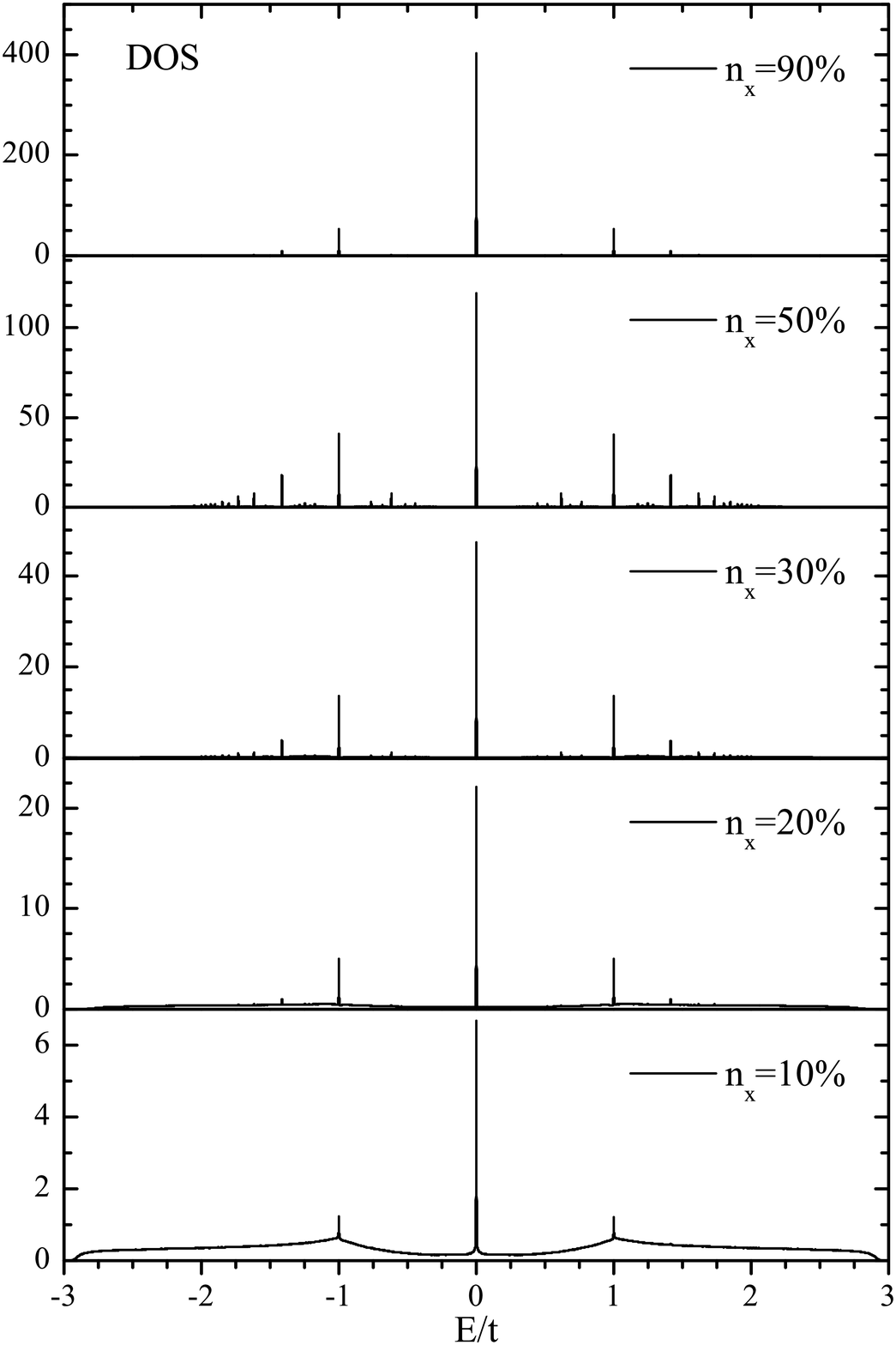}
} \mbox{
\includegraphics[width=8cm]{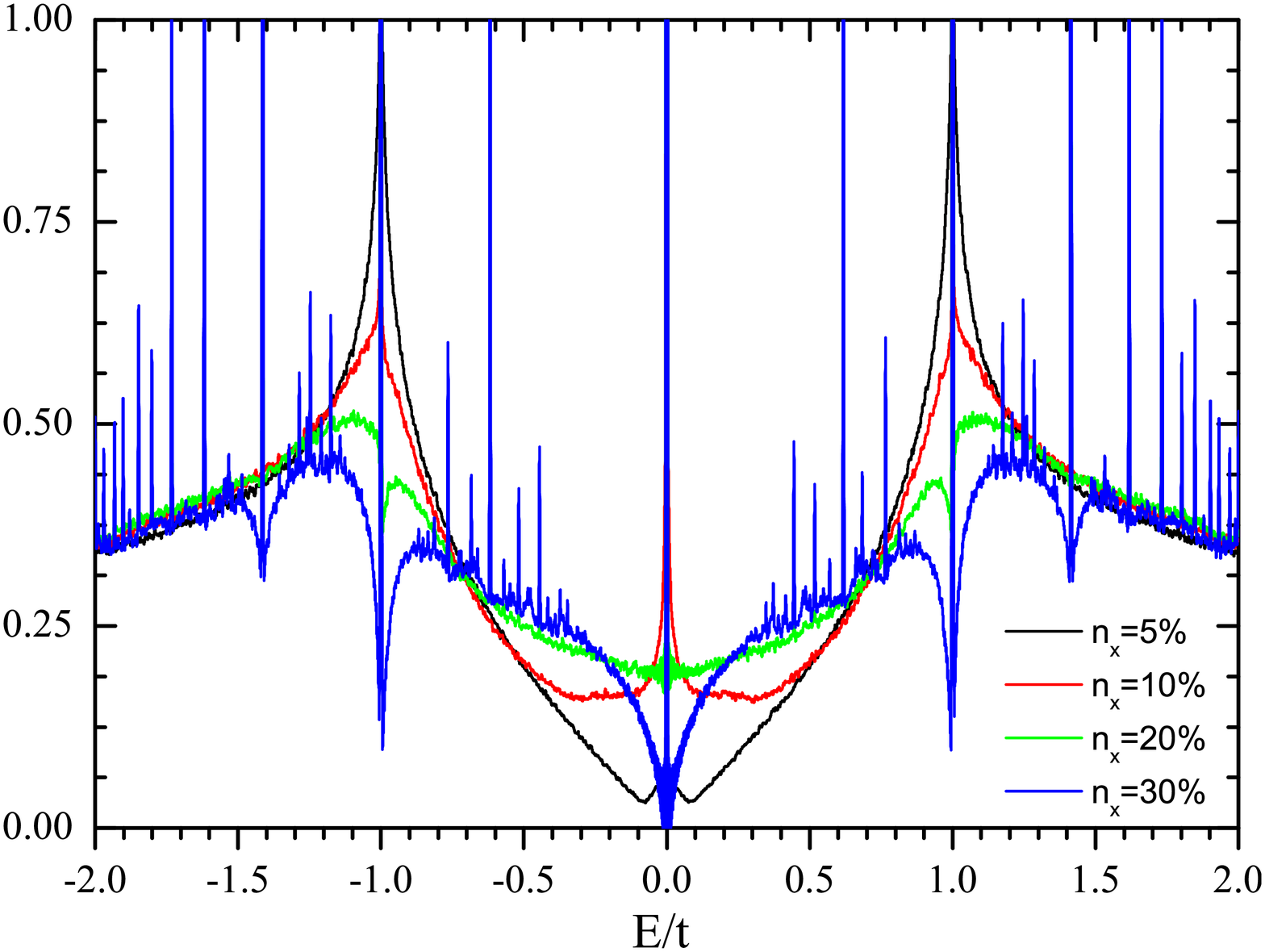}
}
\end{center}
\caption{Density of states (in units of $1/t$) as a function of energy $E$ (in units of $t$) for
the vacancies with large concentrations: $n_{x}=5\%,$ $10\%,$ $20\%,$ $30\%,$
$50\%,$ $90\%$. Sample size is $4096\times 4096$ for $n_{x}\leq 50\%$ and $%
8192\times 8192$ for $n_{x}=90\%$.}
\label{doslargex}
\end{figure}

\begin{figure}[t]
\begin{center}
\includegraphics[width=8cm]{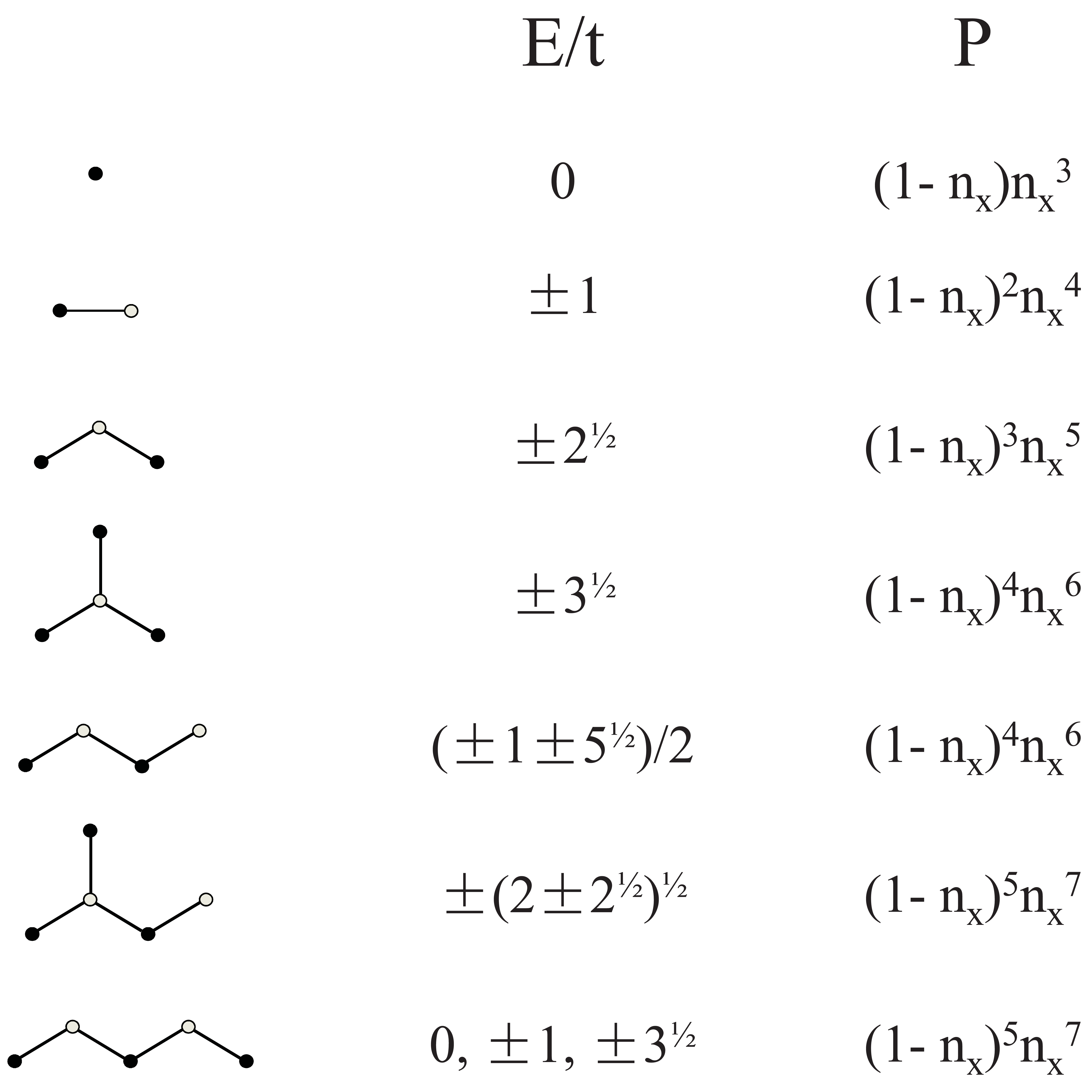}
\end{center}
\caption{Typical atomic structures of most favourable isolated carbon groups
in graphene with large concentration of vacancies. The energy eigenvalues of
each group are listed in the central column (in units of $t$), and $P$ is
the probability of a particular group to be found in a graphene sample.}
\label{carbongroup}
\end{figure}

\begin{figure}[t]
\begin{center}
\includegraphics[width=8cm]{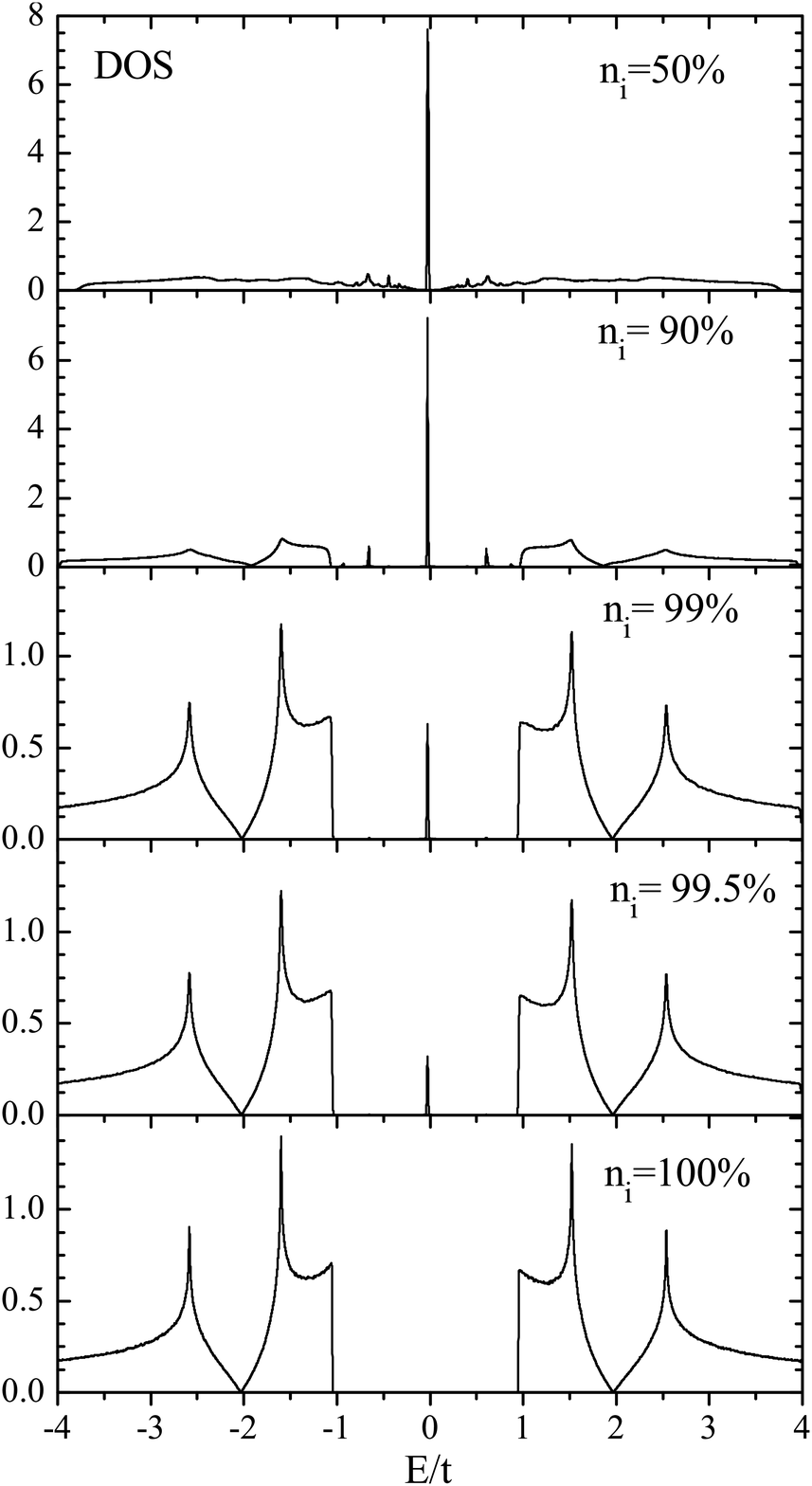}
\end{center}
\caption{Density of states (in units of $1/t$) as a function of energy $E$ (in units of $t$) for
the resonant impurities ($\protect\varepsilon _{d}=-t/16,$ $V=2t$) with
large concentrations: $n_{i}=50\%,$ $90\%,$ $99\%,$ $99.5\%,$ $100\%$.
Sample size is $2048\times 2048.$}
\label{doslargximp}
\end{figure}

\begin{figure}[t]
\begin{center}
\includegraphics[width=8cm]{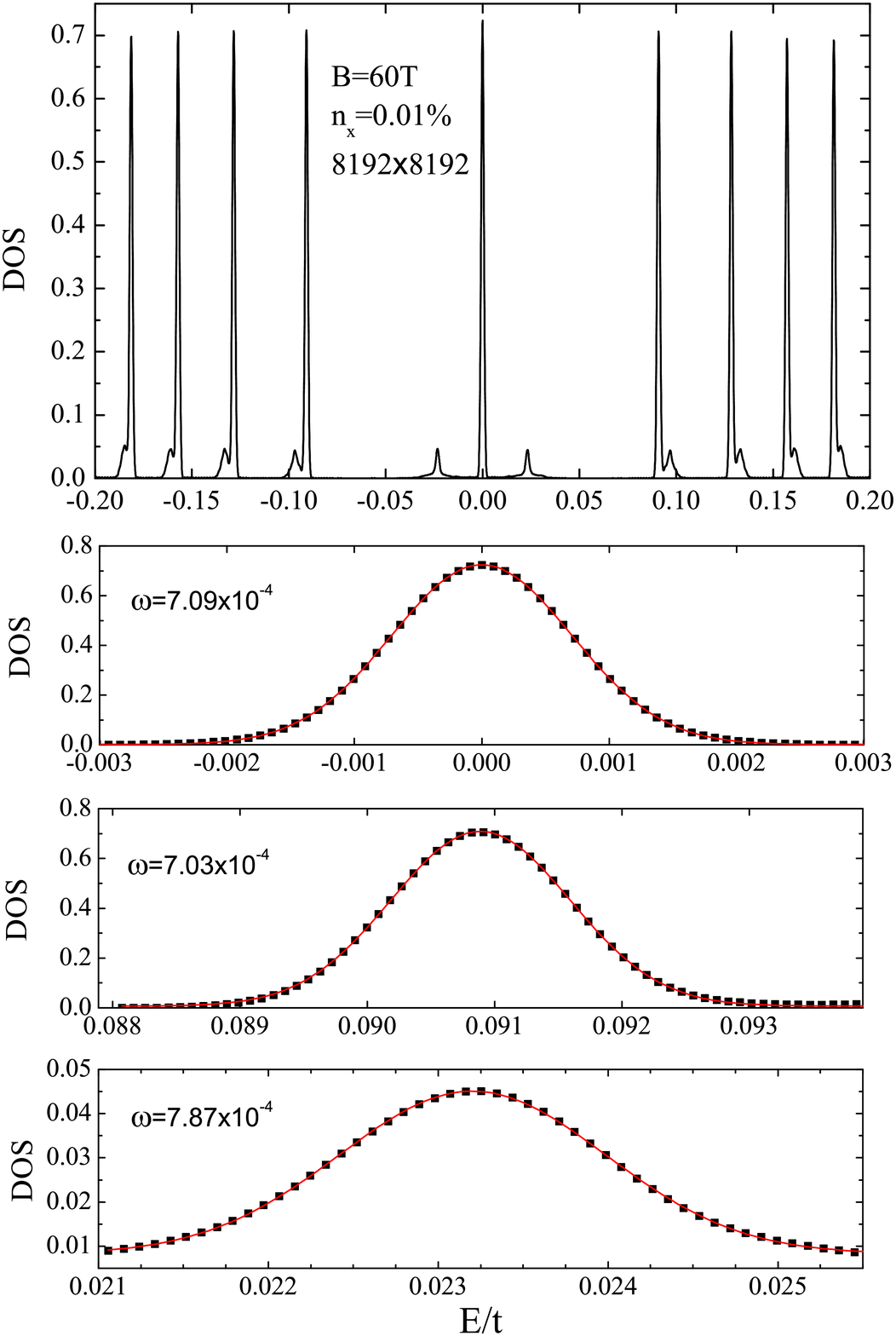}
\end{center}
\caption{(Color online) Density of states (in units of $1/t$) as a function of energy $E$ (in
units of $t$) in the presence of a uniform perpendicular magnetic field ($%
B=60T$) with vacancy concentration $n_{x}=0.01\%$. The red curves are
Gaussian fits of Eq. (\protect\ref{dosgaussian}) centered about each Landau
levels, with $w=7.09\times 10^{-4}$ for $E_{N}=0$ $(N=0)$, $w=7.03\times
10^{-4}$ for $E_{N}=0.0909t$ $(N=1)$, and $w=7.87\times 10^{-4}$ for $%
E=0.0232t$ (between zero and first Landau levels). Sample size is $%
8192\times 8192.$}
\label{dosb60x}
\end{figure}

\begin{figure}[t]
\begin{center}
\includegraphics[width=8cm]{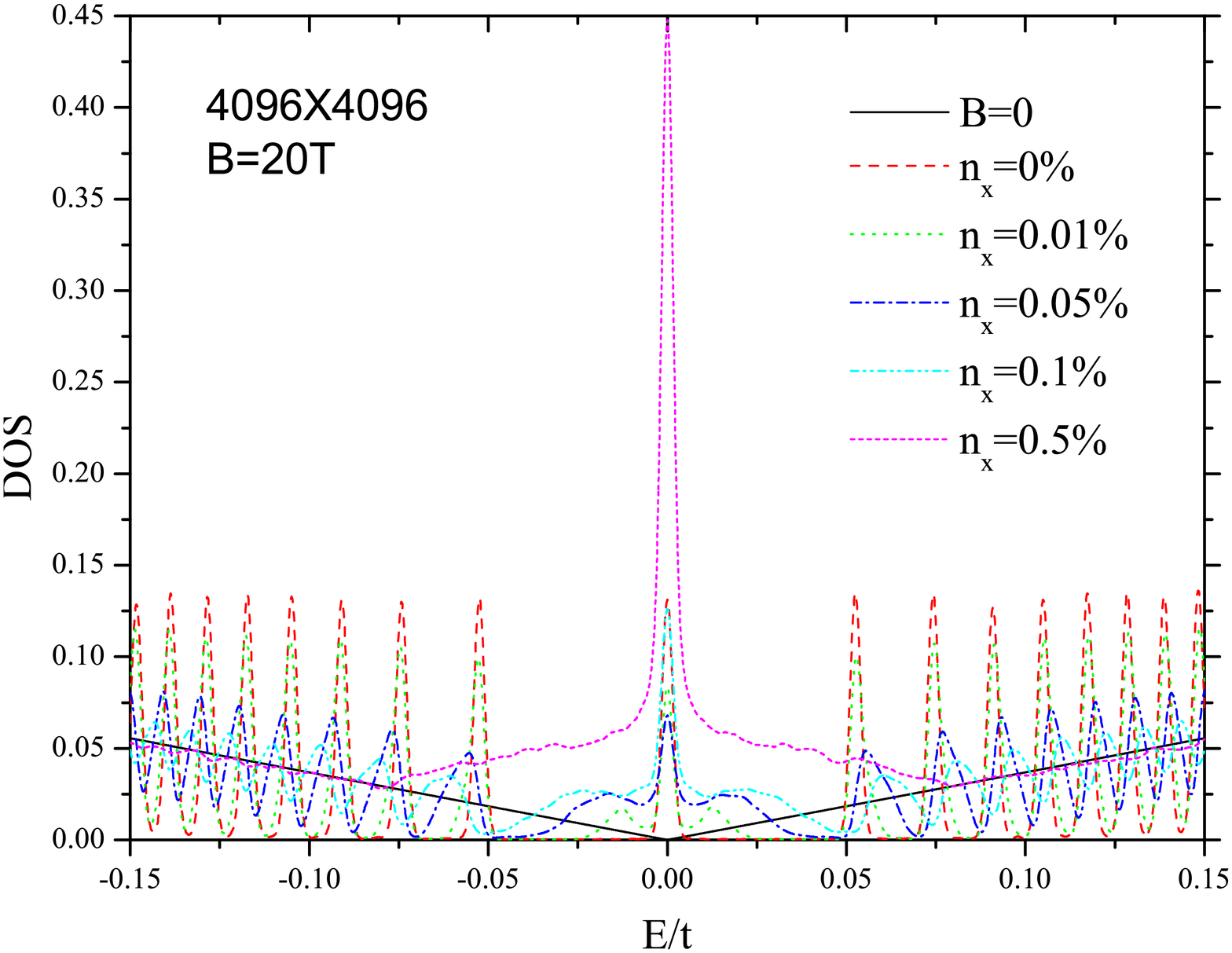}
\end{center}
\caption{(Color online) Density of states (in units of $1/t$) as a function of energy $E$ (in
units of $t$) in the presence of a uniform perpendicular magnetic field ($%
B=20T$) with different vacancy concentrations: $n_{x}=0\%,$ $0.01\%,$ $%
0.05\%,$ $0.1\%,$ $0.5\%$. Sample size is $4096\times 4096.$}
\label{dosbx}
\end{figure}

\begin{figure}[t]
\begin{center}
\includegraphics[width=8cm]{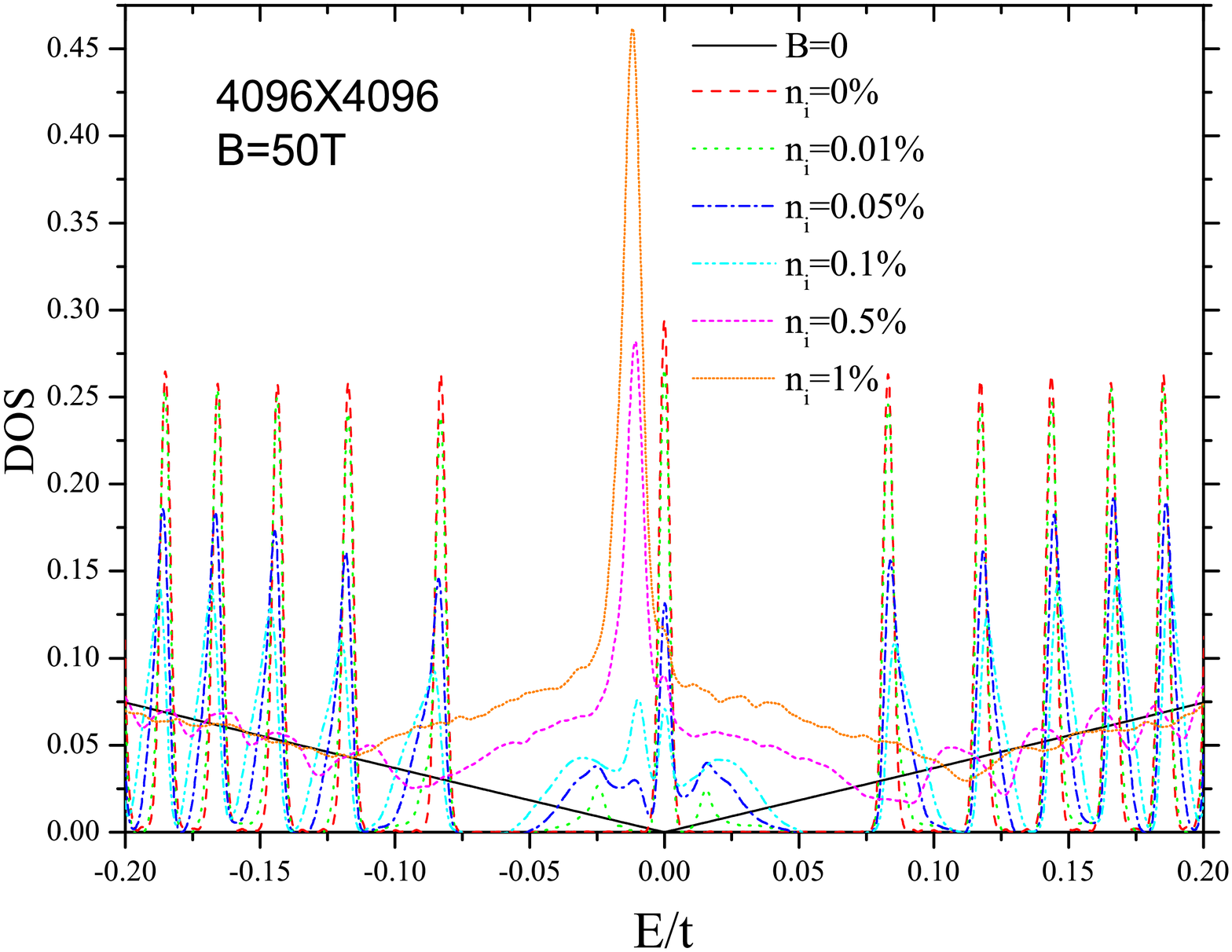}
\end{center}
\caption{(Color online) Density of states (in units of $1/t$) as a function of energy $E$ (in
units of $t$) in the presence of a uniform perpendicular magnetic field ($%
B=50T$) with different hydrogen concentrations $n_{i}=0\%,0.01\%,0.05\%,0.1%
\%,0.5\%,1\%$. Sample size is $4096\times 4096.$}
\label{dosb50ximp}
\end{figure}

The density of states describes the number of states at each energy level.
An algorithm based on the evolution of time-dependent Schr\"{o}dinger
equation (TDSE) to find the eigenvalue distribution of very large matrices
was described in Ref. \onlinecite{Hams2000}. The main idea is to use a
random superposition of all basis states as an initial state $\left\vert
\varphi \left( 0\right) \right\rangle $:%
\begin{equation}
\left\vert \varphi \left( 0\right) \right\rangle =\sum_{i}a_{i}\left\vert
i\right\rangle ,
\end{equation}%
where $\{\left\vert i\right\rangle \}$ are the basis states and $\{a_{i}\}$
are random complex numbers, solve the TDSE at equal time intervals,
calculate the correlation function 
\begin{equation}
\left\langle \varphi \left( 0\right) \right\vert e^{-iHt}\left\vert \varphi
\left( 0\right) \right\rangle ,
\end{equation}%
for each time step (we use units with $\hbar =1$): and then apply the
Fourier transform to these correlation functions to get the local density of
states (LDOS) on the initial state: 
\begin{equation}
d\left( \varepsilon \right) =\frac{1}{2\pi }\int_{-\infty }^{\infty
}e^{i\varepsilon t}\left\langle \varphi \left( 0\right) \right\vert
e^{-iHt}\left\vert \varphi \left( 0\right) \right\rangle dt.  \label{dosf}
\end{equation}%
In practice the Fourier transform in Eq. (\ref{dosf}) is performed by fast
Fourier transformation (FFT). We use a Gaussian window to alleviate the
effects of the finite time used in the numerical time integration of the
TDSE. The number of time integration steps determines the energy resolution:
Distinct eigenvalues that differ more than this resolution appear as
separate peaks in the spectrum. If the eigenvalue is isolated from the rest
of the spectrum, the width of the peak is determined by the number of time
integration steps.

By averaging over different samples (random initial states) we obtain the
density of states: 
\begin{equation}
D\left( \varepsilon \right) =\lim_{S\rightarrow \infty }\frac{1}{S}%
\sum_{p=1}^{S}d_{p}\left( \varepsilon \right) .
\end{equation}%
For a large enough system, for example, graphene crystallite consisting of $%
4096\times 4096\approx 1.6\times 10^{7}$ atoms, one initial random
superposition state (RSS) is already sufficient to contain all the
eigenstates, thus, its LDOS is approximately equal to the DOS of an infinite
system, i.e.,%
\begin{equation}
D\left( \varepsilon \right) \approx d\left( \varepsilon \right) .
\end{equation}%
For the proof of this results and a detailed analysis of this method we
refer to Ref. \onlinecite{Hams2000}. To validate the method, we will compare
the analytical and numerical results for clean graphene.

The numerical solution of the TDSE is carried out by using the Chebyshev
polynomial algorithm, which is based on the polynomial representation of the
operator $U\left( t\right) =e^{-itH}$ (see Appendix A). The Chebyshev
polynomial algorithm is very efficient for the simulation of quantum systems
and conserves the energy of the whole system to machine precision. In order
to reduce the effects of the graphene edges on the electronic properties
(see, e.g., Ref. \onlinecite{DeRaedt2008}), we use periodic boundary
conditions for all the numerical results presented in this paper.

\subsection{DOS of Clean Graphene}

The analytical expression of the density of states of a clean graphene
(ignoring the next-nearest neighbor interaction $t^{\prime }$ and the
on-site energy) was given in Ref.~\onlinecite{Hobson1953} as 
\begin{equation}
\rho \left( E\right) ={\Huge \{}%
\begin{array}{c}
\frac{2E}{t^{2}\pi ^{2}}\frac{1}{\sqrt{F\left( E/t\right) }}\mathbf{K}\left( 
\frac{4E/t}{F\left( E/t\right) }\right) ,0<E<t, \\ 
\frac{2E}{t^{2}\pi ^{2}}\frac{1}{\sqrt{4E/t}}\mathbf{K}\left( \frac{F\left(
E/t\right) }{4E/t}\right) ,t<E<3t,%
\end{array}
\label{dosclean}
\end{equation}%
where $F\left( x\right) $ is given by%
\begin{equation}
F\left( x\right) =\left( 1+x\right) ^{2}-\frac{\left( x^{2}-1\right) ^{2}}{4}%
,
\end{equation}%
and $\mathbf{K}\left( m\right) $ is the elliptic integrals of first kind:%
\begin{equation}
\mathbf{K}\left( m\right) =\int_{0}^{1}dx\left[ \left( 1-x^{2}\right) \left(
1-mx^{2}\right) \right] ^{-1/2}.
\end{equation}

In Fig.~\ref{doscomparesize}, we compare the analytical expression Eq.~(\ref%
{dosclean}) with the numerical results of the density of states for a clean
graphene. One can clearly see that these numerical results fit very well the
analytical expression, and the difference between the numerical and
analytical results becomes smaller when using larger sample size (see the
difference of a sample with $512\times 512$ or $4096\times 4096$ in Fig.~\ref%
{doscomparesize}). In fact, the local density of states of a sample
containing $4096\times 4096$ is approximately the same as the density of
states of infinite clean graphene, which indicates the high accuracy of the
algorithm.

\subsection{DOS of Graphene with Impurities}

Next, we consider the influence of two types of defects on the DOS of
graphene, namely, vacancies and resonant impurities. A vacancy can be
regarded as an atom (lattice point) with and on-site energy $v\rightarrow
\infty $ or with its hopping parameters to other sites being zero. In the
numerical simulation, the simplest way to implement a vacancy it to remove
the atom at the vacancy site. Introducing vacancies in a graphene sheet will
create a zero energy modes (midgap state) \cite%
{Peres2006,Pereira2006,Pereira2008}. The exact analytical wave function
associated with the zero mode induced by a single vacancy in a graphene
sheet was obtained in Ref.\onlinecite{Pereira2006}, showing a quasilocalized
character with the amplitude of the wave function decaying as inverse
distance to the vacancy. Graphene with a finite concentration of vacancies
was studied numerically in Ref. \onlinecite{Pereira2008}. The number of the
midgap states increases with the concentration of the vacancies. The
inclusion of vacancies brings an increase of spectral weight to the
surrounding of the Dirac point ($E=0)$ and smears the van Hove singularities 
\cite{Peres2006,Pereira2008}. Our numerical results (see Fig.~\ref%
{dosximpandx}) confirm all these findings.

Resonant impurities are introduced by the formation of a chemical bond
between a carbon atom from graphene sheet and a carbon/oxygen/hydrogen atom
from an adsorbed organic molecule (CH$_{3}$, C$_{2}$H$_{5}$, CH$_{2}$OH, as
well as H and OH groups)\cite{Wehling2010}. To be specific, we will call
adsorbates hydrogen atoms but actually, the parameters for organic groups
are almost the same \cite{Wehling2010}. The adsorbates are described by the
Hamiltonian $H_{imp}$ in Eq.~(\ref{Hamiltonian}). The band parameters $%
V\approx 2t$ and $\epsilon _{d}\approx -t/16$ are obtained from the \textit{%
ab initio} density functional theory (DFT) calculations~\cite{Wehling2010}.
As we can see from Fig.~\ref{dosximpandx}, small concentrations of vacancies
or hydrogen impurities have similar effects to the DOS of graphene. Hydrogen
adatoms also lead to zero modes and the quasilocalization of the low-energy
eigenstates, as well as to smearing of the van Hove singularities. The shift
of the central peak of the DOS with respect to the Dirac point in the case
of hydrogen impurities is due to the nonzero (negative) on-site potentials $%
\epsilon _{d}$.

Now we consider the electronic structure of graphene with a higher
concentration of defects. Large concentration of vacancies in graphene leads
to well pronounced symmetric peaks in the DOS: a very high central peak at
the Dirac point, two small peaks at the Van Hove singularities, and tiny
peaks at  $\left\vert E\right\vert /t=0.618,0.766,1.414,1.618,1.732,1.848$ \
(see Fig.~\ref{doslargex}). These results indicate the emergence of small
pieces of isolated carbon groups, shown in Fig.~\ref{carbongroup}. The
positions of the peaks in the DOS match very well with the energy
eigenvalues of these small subgroups. For example, non-interacting carbon
atoms contribute to the peak at Dirac point, and isolated pairs contribute
to the peaks at Van Hove singularities. Graphene with very high vacancy
concentration, e.g., $n_{x}$\ = 90\%, is mainly a sheet of non-interacting
carbon atoms, with small amount of isolated pairs, and tiny amounts of
isolated triples. Only the peaks corresponding to these groups appear in the
calculated DOS of $n_{x}$\ = 90\% in Fig.~\ref{doslargex}.

Graphene with $100\%$ concentration of hydrogen impurities is not graphene,
but pure graphane~\cite{Elias2009}. Graphane is shown to be an insulator
because of the existence of a band gap (in our model, $2t$), see the bottom
panel in Fig.~\ref{doslargximp}. Graphene with large concentration ($n_{i}$)
of hydrogen impurities corresponds to graphane with small concentrations ($%
1-n_{i}$) of vacancies of hydrogen atoms, which leads, again, to appearance
of localized midgap states (shifted from zero due to nonzero $\epsilon _{d}$%
) on the carbon atoms which have no hopping integrals to any hydrogen, see
these central peaks in Fig.~\ref{doslargximp}. Despite the fact that our
model is oversimplified for dealing with finite concentrations of hydrogen
(in general, parameters of impurities should be concentration dependent,
direct hopping between hydrogens should be taken into account, etc.), this
conclusion is in an agreement with first principle calculations \cite%
{Lebegue2009}.

\subsection{DOS of Graphene with Impurities in the Magnetic Field}

A magnetic field perpendicular to a graphene layer leads to discrete Landau
energy levels. The energy of the Landau levels of clean graphene is given by 
\cite{r2,r3} 
\begin{equation}
E_{N}=sgn(N)\sqrt{2e\hbar v_{F}^{2}B\left\vert N\right\vert },
\label{Landaulevel}
\end{equation}%
where in the nearest-neighbor tight binding model 
\begin{equation}
v_{F}/t=3a/2\hbar .
\end{equation}%
Our numerical calculations reproduces the positions of the Landau levels.
Introducing impurities or disorders in graphene will broaden the Landau
levels. Fig.~\ref{dosb60x} presents the numerical results for a uniform
perpendicular magnetic field ($B=60T$) applied to a $8192\times 8192$
graphene sample with a small concentration of vacancies ($n_{x}=0.01\%$).
The spectral distribution near each Landau level fits well to the Gaussian
function 
\begin{equation}
\rho \left( E\right) =A\exp \left[ -\frac{\left( E-E_{N}\right) ^{2}}{2w^{2}}%
\right] ,  \label{dosgaussian}
\end{equation}%
with $w\approx 7\times 10^{-4}t$. Between two Laudau levels, there are extra
peaks which also fit to a Gaussian distribution with $w\approx 8\times
10^{-4}t$. These additional localized states were also found in other
numerical simulations~\cite{Pereira2008b} of much smaller $96\times 60$
samples with a stronger magnetic field ($B\approx 400T$) and larger
concentration of vacancies ($n_{x}=0.21\%$ and $0.42\%$).

Increasing the concentration of the vacancies will smear and suppress the
Landau levels except the one at zero energy\cite{Peres2006}, see Fig.~\ref%
{dosbx}. The zero-energy Landau level seems to be robust with respect to
resonant impurities since the latter form their own midgap states.

The presence of hydrogen impurities has similar effects on the spectrum as
in the case of vacancies (compare Fig. \ref{dosbx} and~\ref{dosb50ximp})
except that, because of the non-zero on-site energy ($\epsilon _{d}$) of
hydrogen sites, the zero-energy Landau level splits into two for a certain
range of hydrogen concentrations (for example, see $n_{i}=0.05\%$ in Fig.~%
\ref{dosb50ximp}). The peak at the neutrality point corresponds to the
original zero-energy Landau level whereas the other one originates mainly
from hybridization with hydrogen atoms. The splitting of zero-energy Landau
level by other kinds of disorder is also observed, for example, with random
nearest-neighbor hopping between carbon atoms as reported in Ref. %
\onlinecite{Pereira2009}.

For small concentration of hydrogen impurities ($n_{i}=0.01\%$ in Fig.~\ref%
{dosb50ximp}), %),
there are also extra peaks between zero and first Landau levels, similar as
in the case for low concentration of vacancies. The difference is that these
two extra peaks are not symmetric around the neutrality point, because of
non-zero on-site energy ($\epsilon _{d}$).

\section{Quasieigenstates}

\begin{figure}[t]
\begin{center}
\mbox{
\includegraphics[width=4cm]{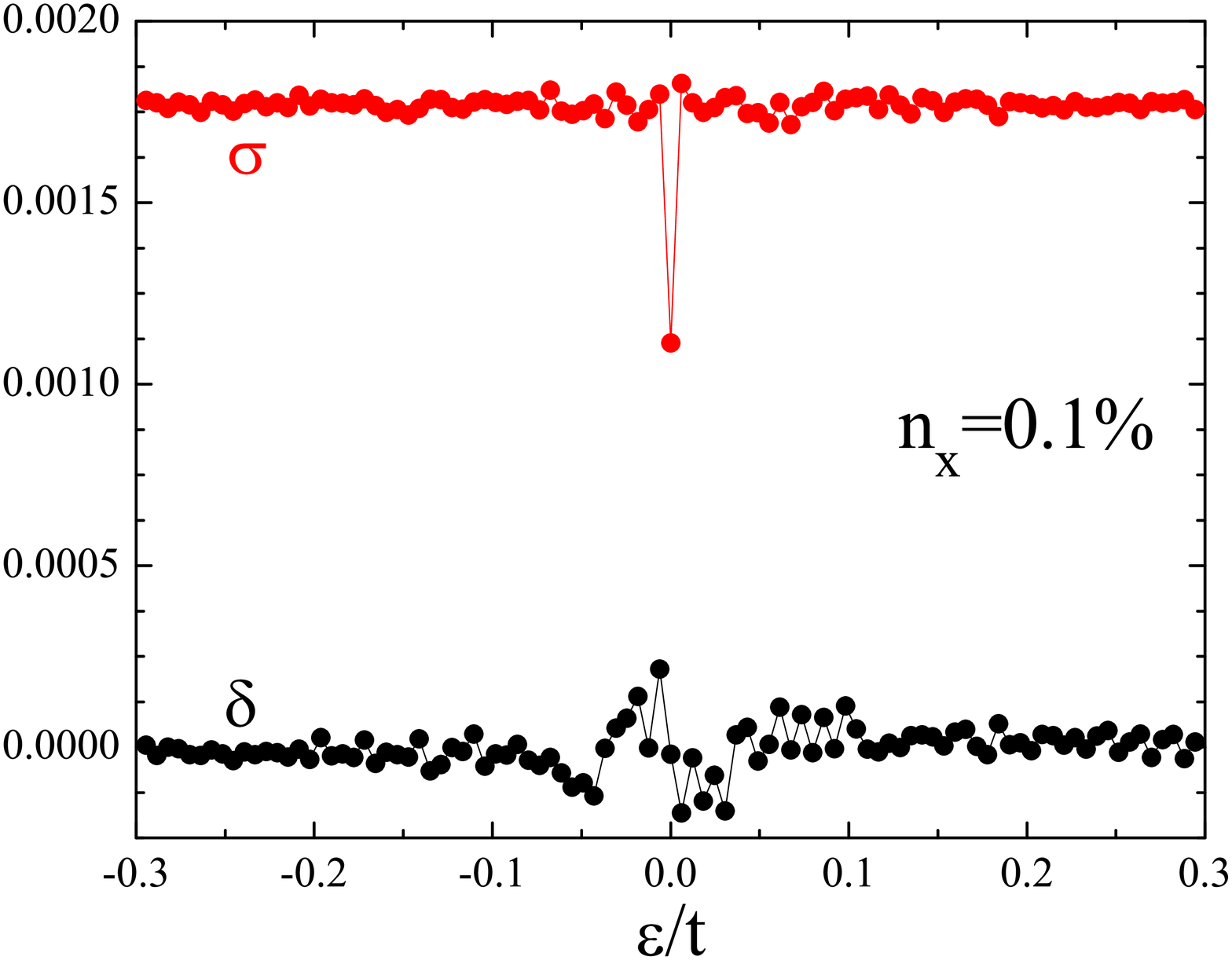}
\includegraphics[width=4cm]{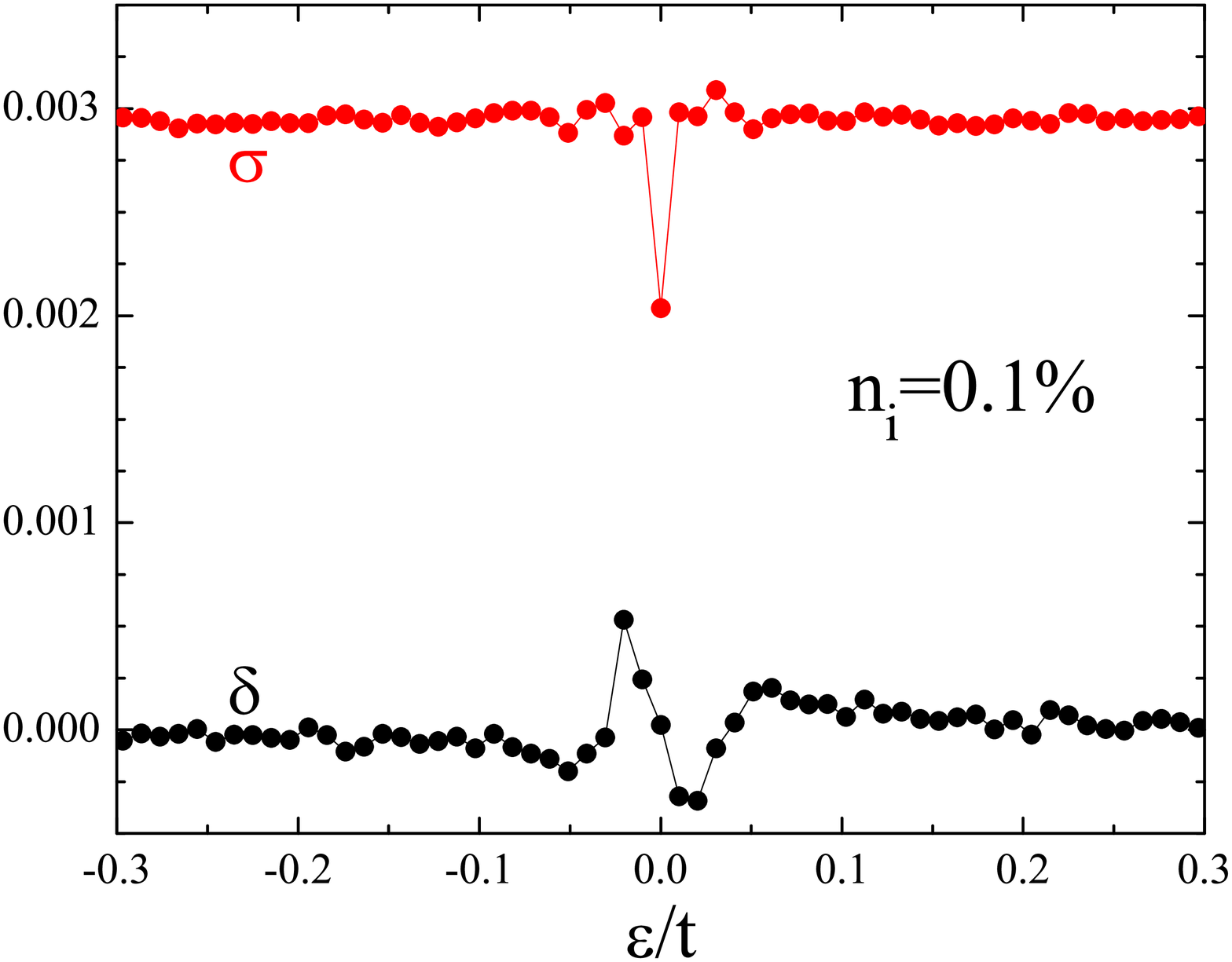}
}
\end{center}
\caption{(Color online) The error ($\protect\delta $ and $\protect\sigma $)
of the approximation of $\left\vert \Psi \left( \protect\varepsilon \right)
\right\rangle $ of a quasieigenstate in a graphene sample ($4096\times 4096$%
) with vacancies or hydrogen (($\protect\varepsilon _{d}=-t/16,$ $V=2t$)
impurities. The concentration of the defeats is $0.1\%$.}
\label{error}
\end{figure}

\begin{figure*}[t]
\begin{center}
\mbox{
\includegraphics[width=8cm]{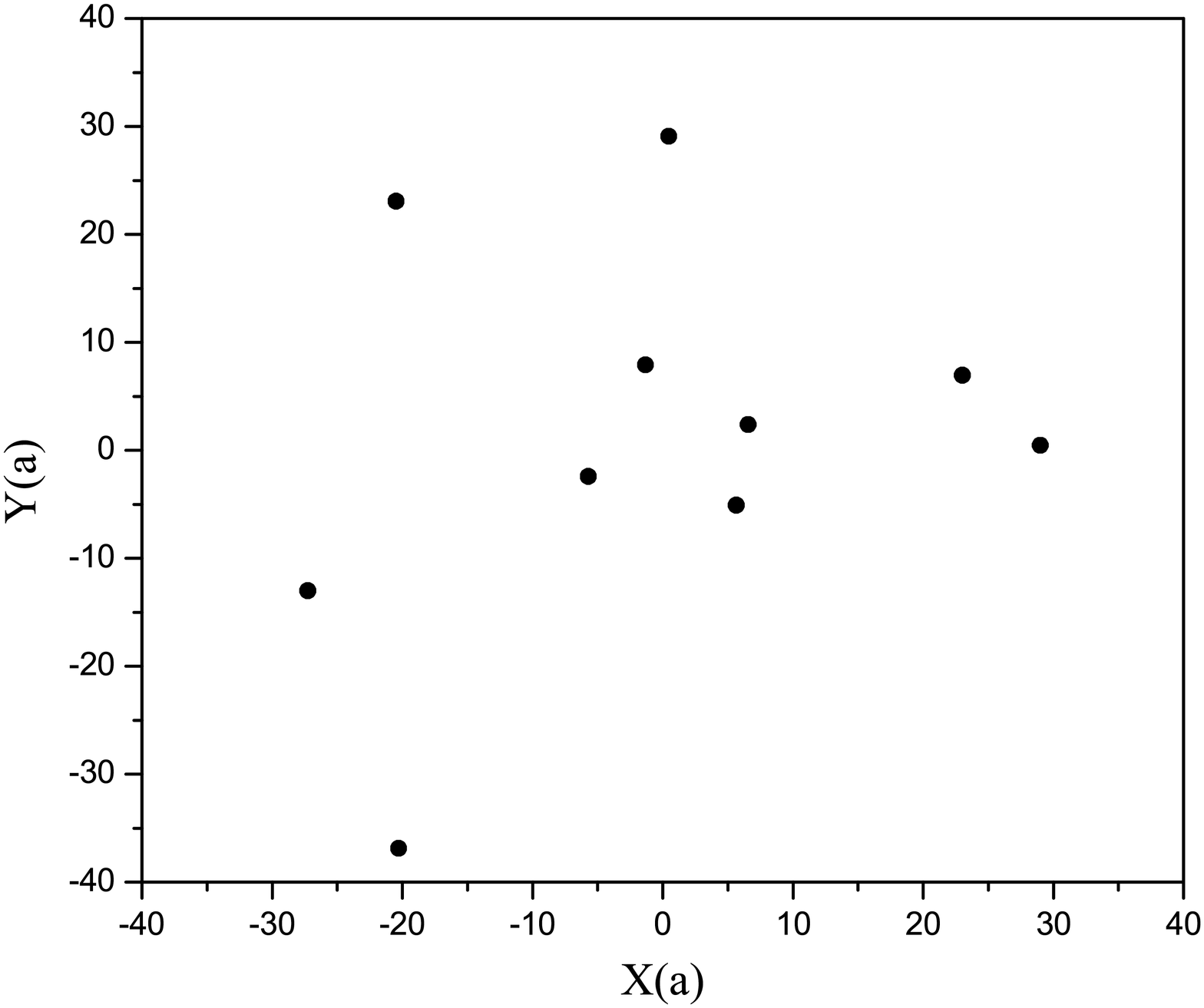}
\includegraphics[width=8cm]{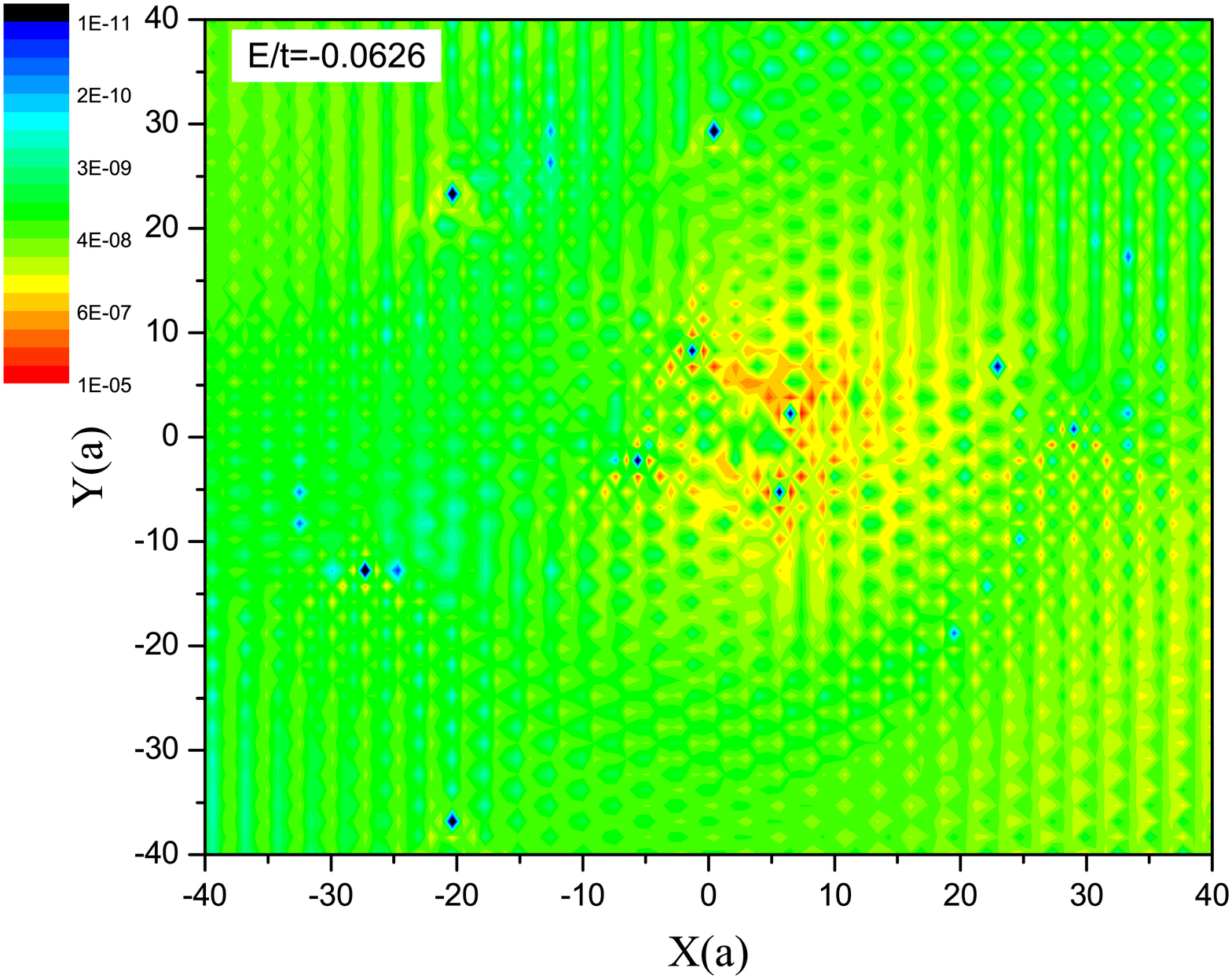}
} 
\mbox{
\includegraphics[width=8cm]{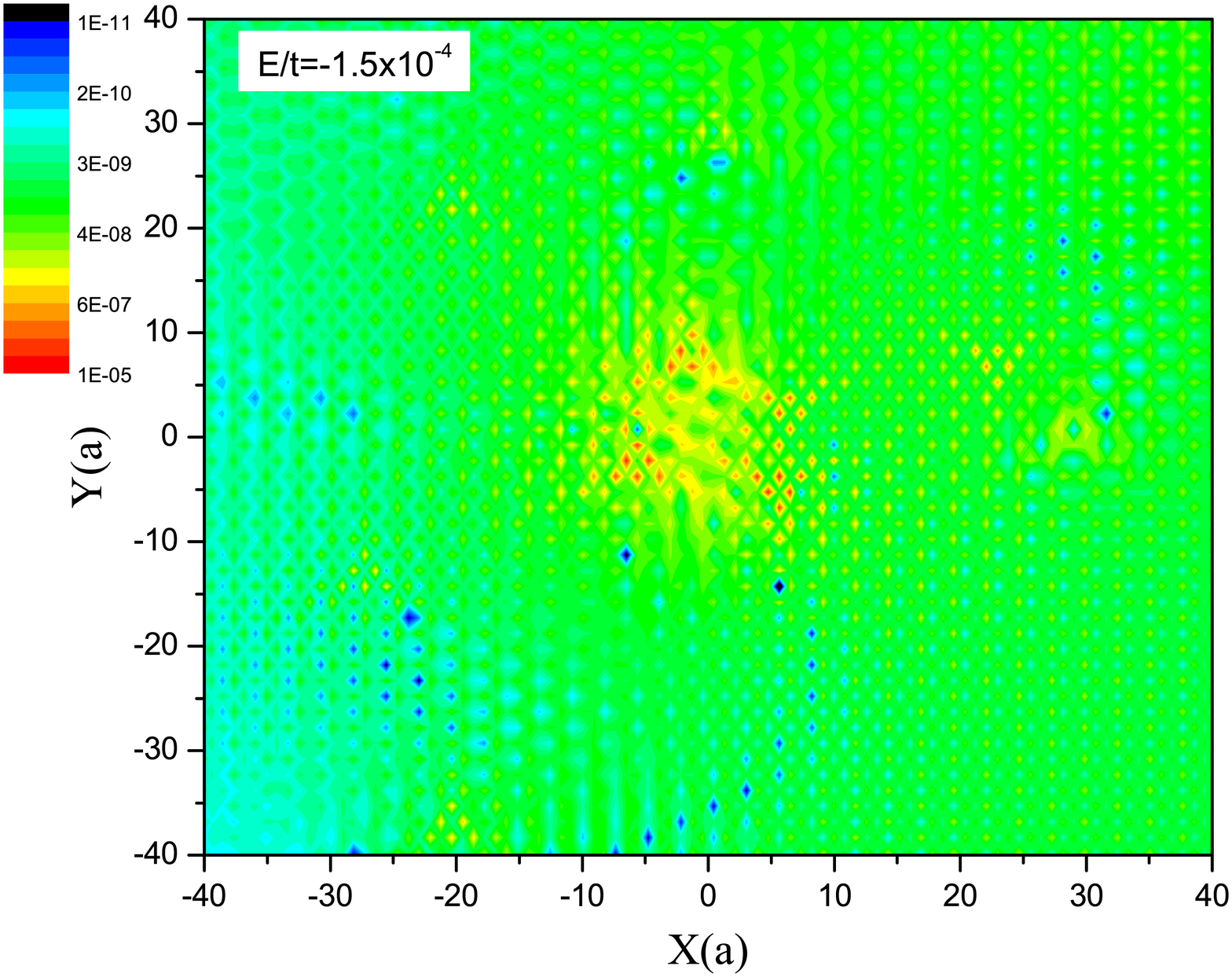}
\includegraphics[width=8cm]{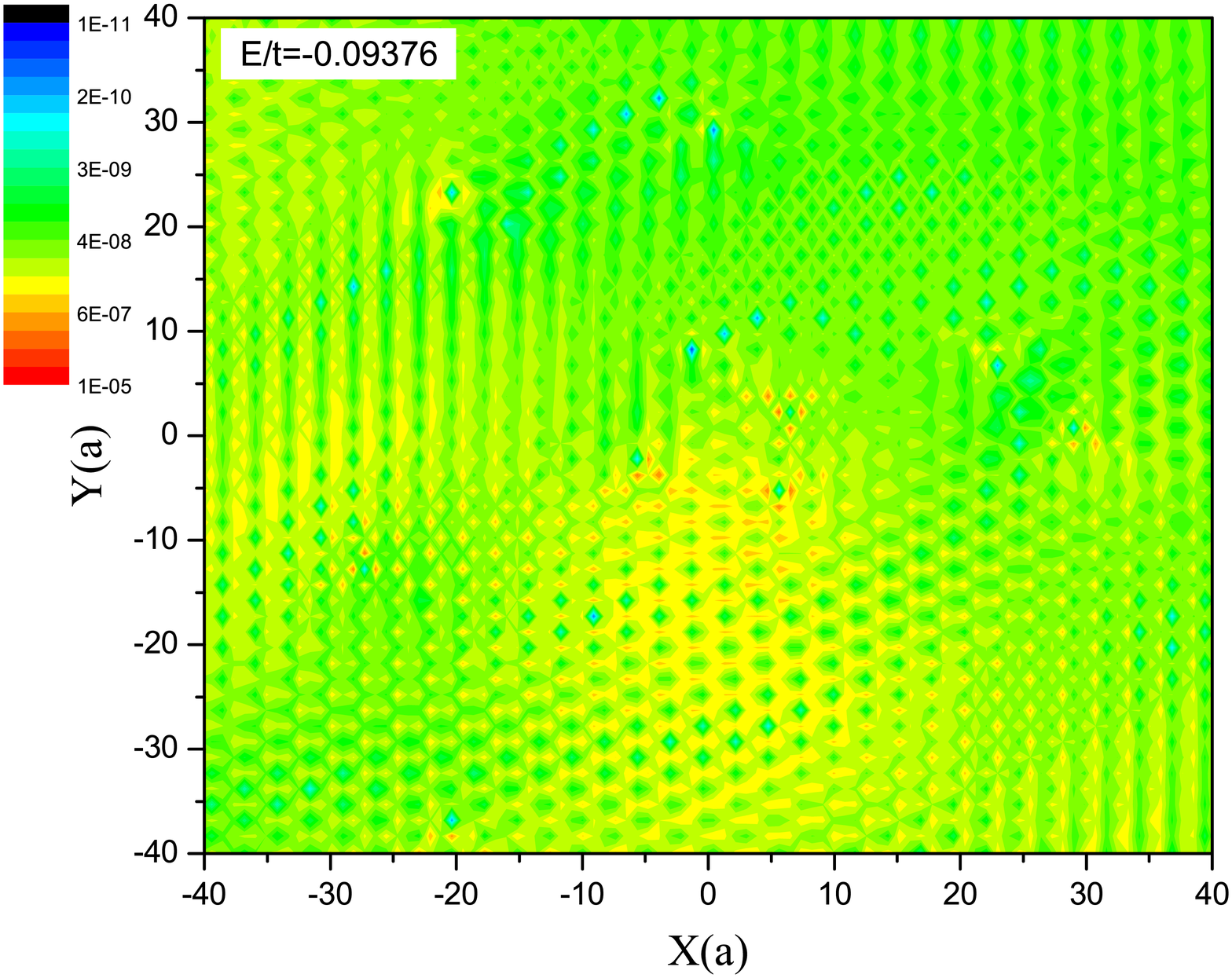}
} 
\mbox{
\includegraphics[width=8cm]{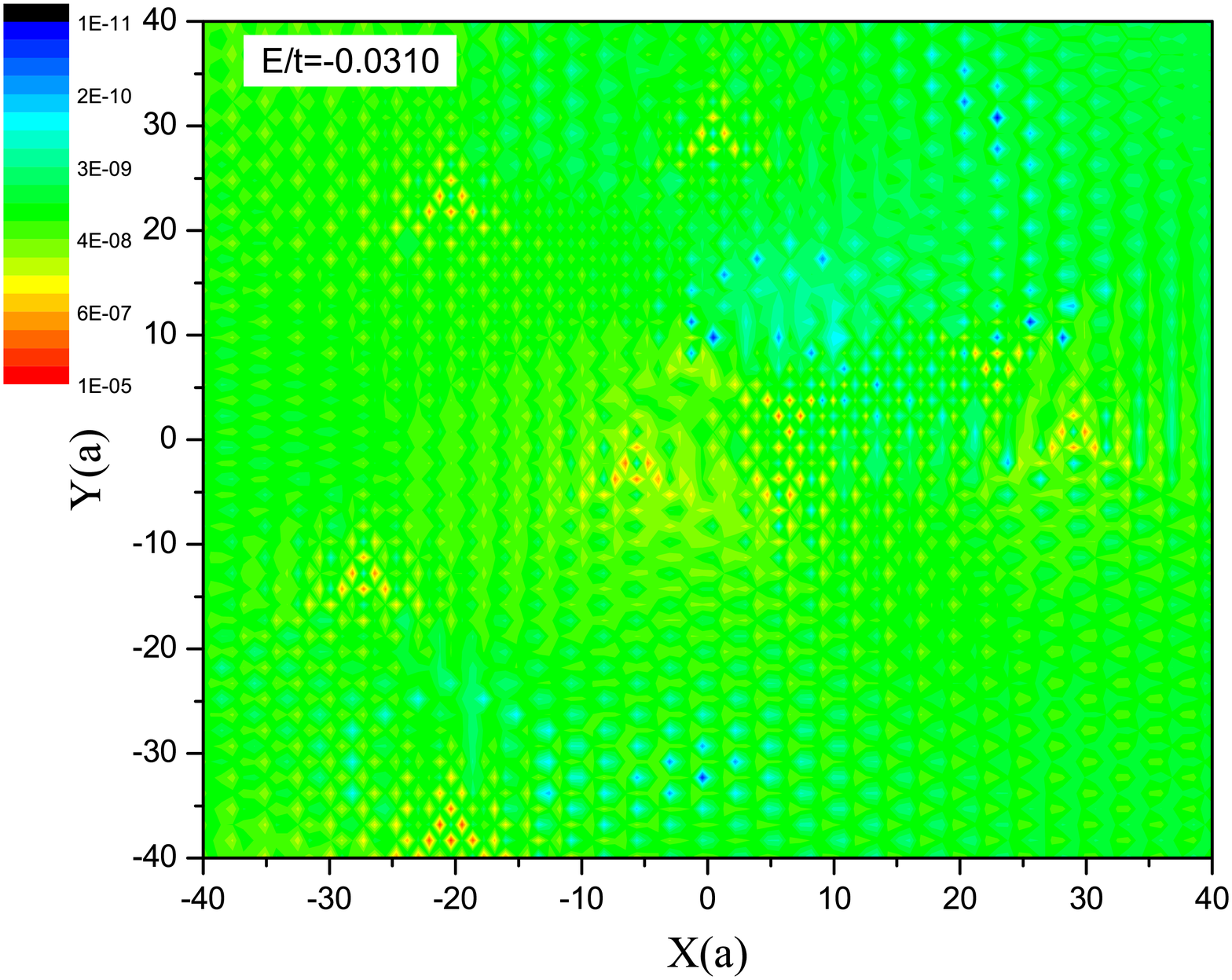}
\includegraphics[width=8cm]{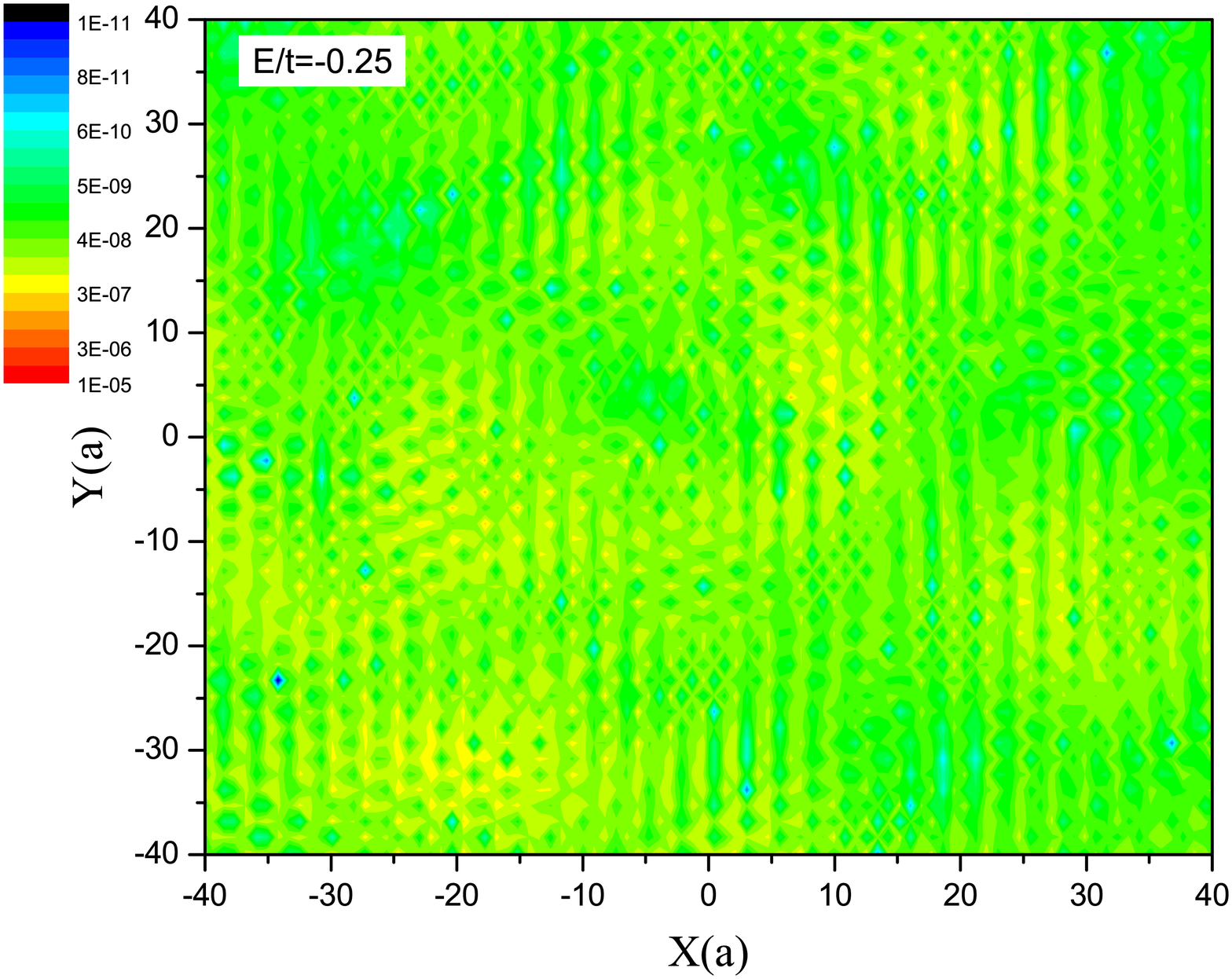}
}
\end{center}
\caption{(Color online) Position of hydrogen impurities (black dots in the
top left panel) and contour plot of the amplitudes of the quasieigenstates
in the central part of a graphene sample ($4096\times 4096$) with different
energies. The concentration of the hydrogen impurities ($\protect\varepsilon %
_{d}=-t/16,$ $V=2t$) is $0.1\%$.}
\label{contourximp1}
\end{figure*}

\begin{figure*}[t]
\begin{center}
\mbox{
\includegraphics[width=8cm]{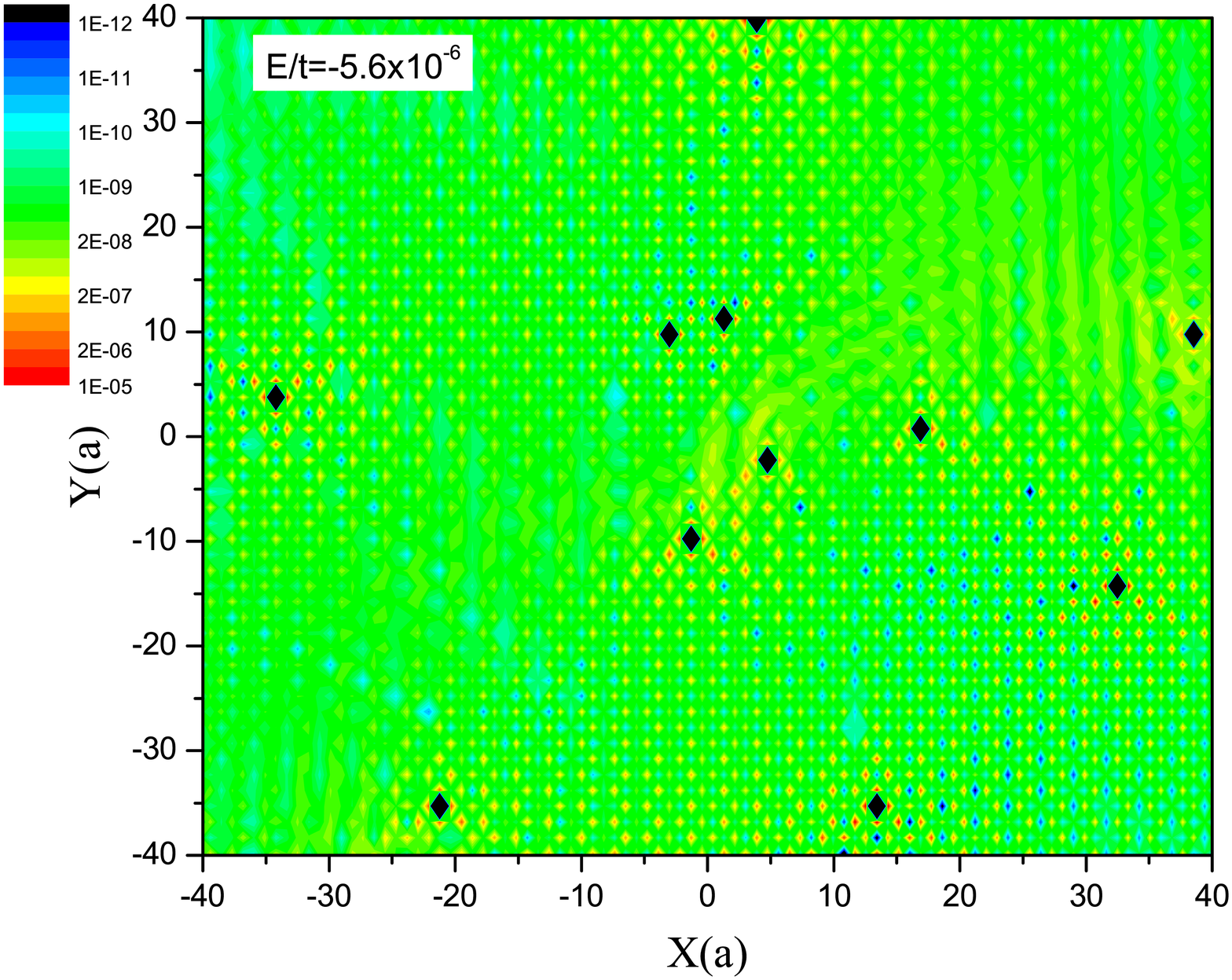}
\includegraphics[width=8cm]{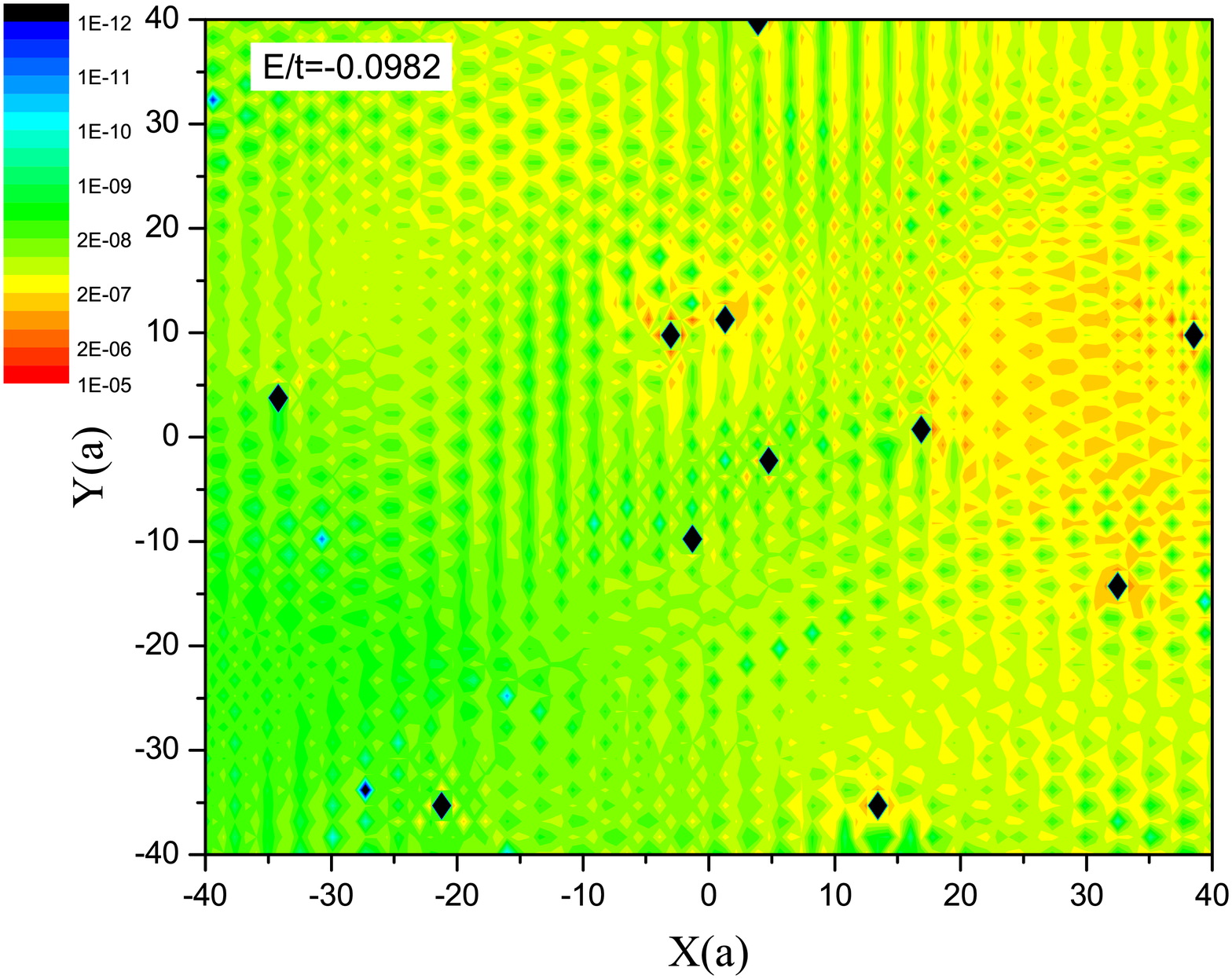}
} 
\mbox{
\includegraphics[width=8cm]{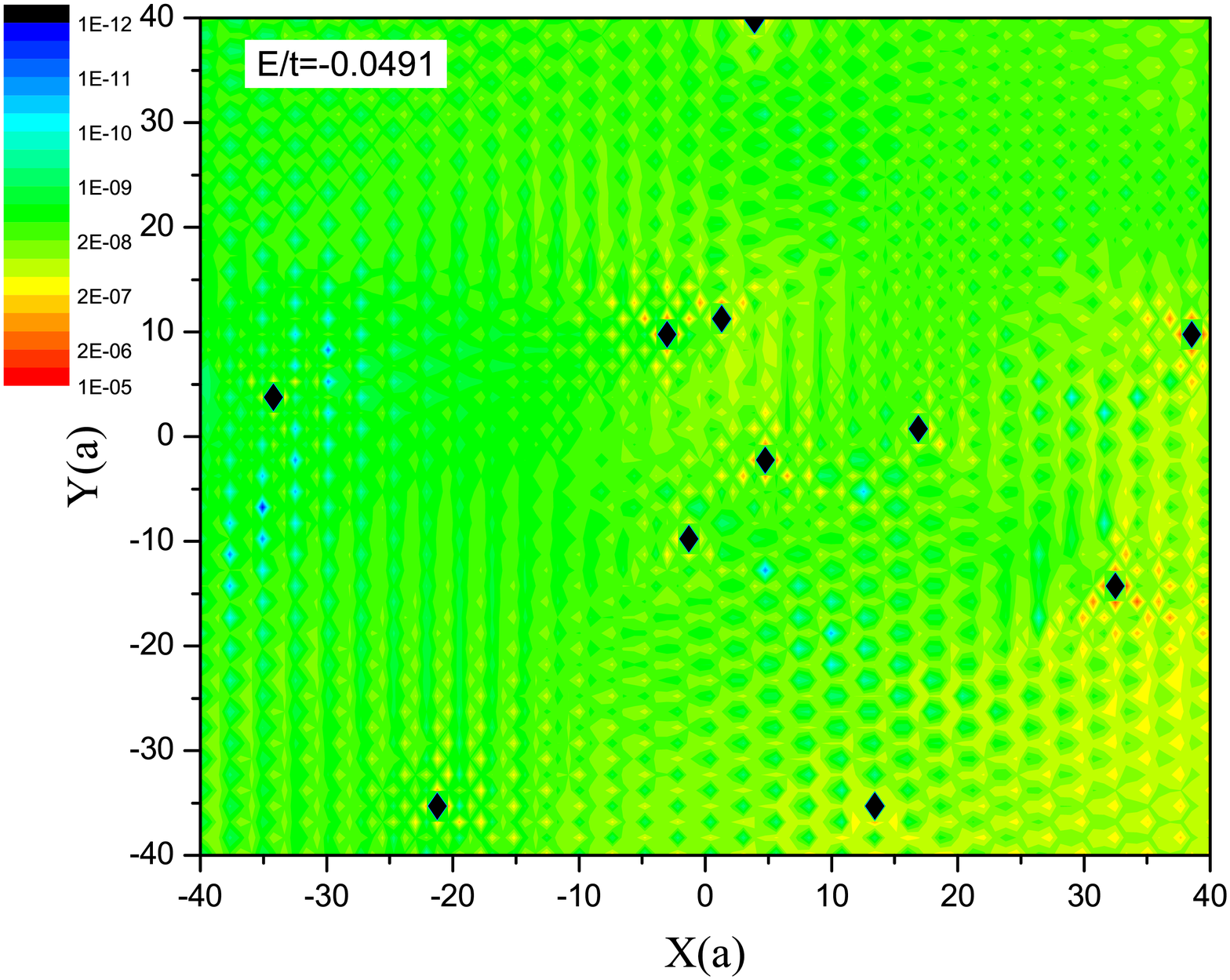}
\includegraphics[width=8cm]{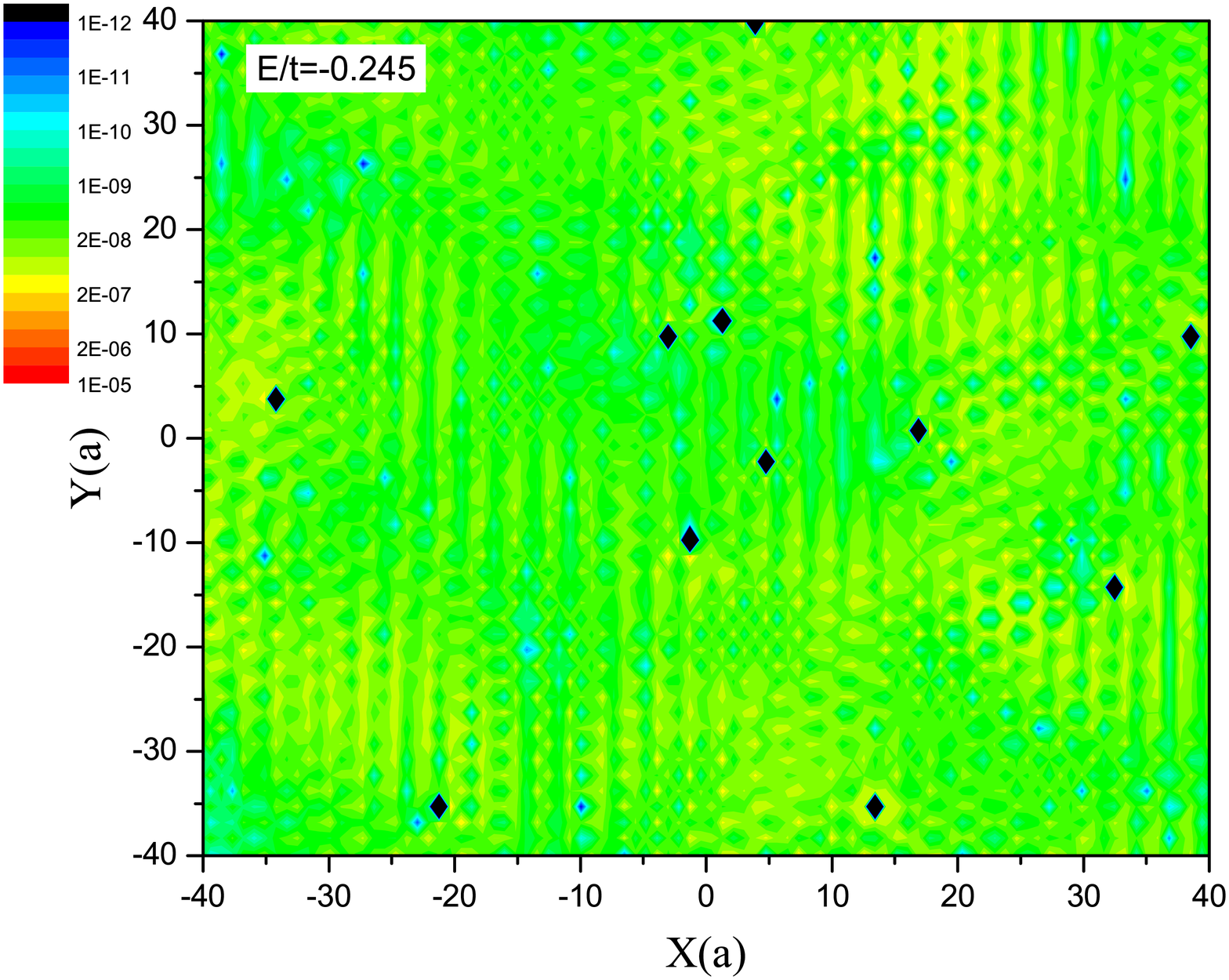}
}
\end{center}
\caption{(Color online) Contour plot of the amplitudes of the
quasieigenstates in the central part of a graphene sample ($4096\times 4096$%
) with different energies. The concentration of the vacancy impurities
(indicated by black dots) is $0.1\%$.}
\label{contourx1}
\end{figure*}

For the general Hamiltonian (\ref{Hamiltonian}) and for samples containing
millions of carbon atoms, in practice, the eigenstates cannot be obtained
directly from matrix diagonalization. An approximation of these eigenstates,
or a superposition of degenerate eigenstates can be obtained by using the
spectrum method \cite{Kosloffa1983}. Let $\left\vert \varphi \left( 0\right)
\right\rangle =\sum_{n}A_{n}\left\vert n\right\rangle $ be the initial state
of the system, and $\{\left\vert n\right\rangle \}$ are the complete set of
energy eigenstates. The state at time $t$ is 
\begin{equation}
\left\vert \varphi \left( t\right) \right\rangle =e^{-iHt}\left\vert \varphi
\left( 0\right) \right\rangle .  \label{phit}
\end{equation}%
Performing the Fourier transform of $\left\vert \varphi \left( t\right)
\right\rangle $ one obtains the expression 
\begin{eqnarray}
\left\vert \widetilde{\Psi }\left( \varepsilon \right) \right\rangle &=&%
\frac{1}{2\pi }\int_{-\infty }^{\infty }dte^{i\varepsilon t}\left\vert
\varphi \left( t\right) \right\rangle  \notag  \label{Eigenstate1} \\
&=&\frac{1}{2\pi }\sum_{n}A_{n}\int_{-\infty }^{\infty }dte^{i\left(
\varepsilon -E_{n}\right) t}\left\vert n\right\rangle  \notag \\
&=&\sum_{n}A_{n}\delta \left( \varepsilon -E_{n}\right) \left\vert
n\right\rangle ,
\end{eqnarray}%
which can be normalized as 
\begin{eqnarray}
\left\vert \Psi \left( \varepsilon \right) \right\rangle &=&\frac{1}{\sqrt{%
\sum_{n}\left\vert A_{n}\right\vert ^{2}\delta \left( \varepsilon
-E_{n}\right) }}\sum_{n}A_{n}\delta \left( \varepsilon -E_{n}\right)
\left\vert n\right\rangle .  \notag \\
&&
\end{eqnarray}%
It is clear that $\left\vert \Psi \left( \varepsilon \right) \right\rangle $
is an eigenstate if it is a single (non-degenerate) state, and some
superposition of degenerate eigenstates with the energy $\varepsilon $,
otherwise.

In general, $\left\vert \Psi \left( \varepsilon \right) \right\rangle $ will
not be an eigenstate but may be close to one and therefore we call it 
\textit{quasieigenstate}. Although $\left\vert \Psi \left( \varepsilon
\right) \right\rangle $ is written in the energy basis, the actual basis
used to represent the state $\left\vert \varphi \left( t\right)
\right\rangle $ can be any orthogonal and complete basis. It is convenient
to introduce two variables $\delta \left( \varepsilon \right) $ and $\sigma
\left( \varepsilon \right) $ to measure the difference between a true
eigenstate and the quasieigenstate $\left\vert \Psi \left( \varepsilon
\right) \right\rangle $: 
\begin{eqnarray}
\delta \left( \varepsilon \right) &=&\left\langle \Psi \left( \varepsilon
\right) |H|\Psi \left( \varepsilon \right) \right\rangle -\varepsilon , \\
\sigma \left( \varepsilon \right) &=&\sqrt{\left\langle \Psi \left(
\varepsilon \right) |H^{2}|\Psi \left( \varepsilon \right) \right\rangle
-\left\langle \Psi \left( \varepsilon \right) |H|\Psi \left( \varepsilon
\right) \right\rangle ^{2}}.
\end{eqnarray}%
As $\delta \left( \varepsilon \right) $ is a measure of the energy shift and 
$\sigma \left( \varepsilon \right) $ is the variance of the approximation,
both variables should be zero if $\left\vert \Psi \left( \varepsilon \right)
\right\rangle $ is a quasieigenstate with the energy $\varepsilon $. From
numerical experiments (results not shown), we have found two ways to improve
the accuracy of the quasieigenstates. One is that the Fourier transform
should be performed on the states from both positive and negative times, and
the other is that the wave function $\left\vert \varphi \left( t\right)
\right\rangle $ should be multiplied by a window function (Hanning window 
\cite{NumericalRecipes}) $(1+\cos (\pi t/T))/2$ before performing the
Fourier transform, $T$ being the final time of the propagation. The
propagation in both positive and negative time is necessary to keep the
original form of the integral in Eq.~(\ref{Eigenstate1}), and the use of a
window improves the approximation to the integrals.

In Fig.~\ref{error} we show $\delta \left( \varepsilon \right) $ and $\sigma
\left( \varepsilon \right) $ of the calculated quasieigenstates in graphene
with vacancy or hydrogen impurities. The time step used in the propagation
of the wave function is $\tau =1$ in the case of vacancies and $\tau =0.6$
in the case of hydrogen impurity. The total number of time steps is $N_{t}=$ 
$2048$ in both cases. One can see that the errors in the energy of $%
\left\vert \Psi \left( \varepsilon \right) \right\rangle $ are quite small ($%
\left\vert \delta \left( \varepsilon \right) \right\vert <5\times 10^{-4}$),
and the standard deviation $\sigma \left( \varepsilon \right) $ is less than 
$2\times 10^{-3}$ and $3\times 10^{-3}$ for vacancies and hydrogen
impurities, respectively. The value $\sigma \left( \varepsilon \right) $ is
smaller in the case of the vacancies due to the larger time step and larger
propagation time used. The fluctuations of $\delta \left( \varepsilon
\right) $ in the region close to the neutrality point ($\varepsilon =0$) are
due to the error introduced by the finite discrete Fourier transform in Eq.~(%
\ref{Eigenstate1}), because near the neutrality point, the finite discrete
Fourier transform may mix components from the eigenstates in the opposite
side of the spectrum. In fact, it would be more accurate to directly use $%
\left\langle \Psi \left( \varepsilon \right) |H|\Psi \left( \varepsilon
\right) \right\rangle $ instead of $\varepsilon $ as the energy of the
quasieigenstate. Notice that the error of $\sigma \left( \varepsilon \right) 
$ with $\varepsilon =0$ is smaller than in the case of nonzero $\varepsilon $%
, since for $\varepsilon =0$ there is no error due the combination of the
factor $e^{i\varepsilon t}(=1)$ with the state $\left\vert \varphi \left(
t\right) \right\rangle $. All the errors of $\delta \left( \varepsilon
\right) $ and $\sigma \left( \varepsilon \right) $ as well as these
fluctuations around $\delta \left( \varepsilon \right) $ can be reduced by
increasing the time step $\tau $ and/or total number of time steps $N_{t}$.

Although quasieigenstates are not exact eigenstates, they can be used to
calculate the electronic properties of the sample, such as the DC
conductivity (as will be shown later). The contour plot of the amplitudes of
the quasieigenstates directly reveals the structure of the eigenstates with
certain eigenenergy, for example, the quasilocalization of the low-energy
states around the vacancy or hydrogen impurity, see Fig.~\ref{contourximp1}
and Fig.~\ref{contourx1}. The quasilocalization of the states around the
impurities occurs not only for zero energy, but also for quasieigenstates
with the energies close to the neutrality point. This quasilocalization
leads to an increase of the spectral weight in the vicinity of the Dirac
point ($E=0$), see Fig.~\ref{dosximpandx}. The states with larger
eigenenergy are extended and robust to small concentration of impurities,
and their spectral weight is close to that in clean graphene. One can see
that in the case of hydrogen impurities, the quasieigenstates that are close
enough to the impurity states, i.e., $E/t=-0.0626\approx \varepsilon _{d}$
in Fig.~\ref{contourximp1}, are distributed in the whole region around
hydrogen atoms. The carbon atoms coupled to hydrogens look like
\textquotedblleft vacancies\textquotedblright, with very small probability
amplitudes, which explains why hydrogen impurities and vacancies produce
similar effects on the electronic properties of graphene.

\section{Optical Conductivity}

\begin{figure*}[t]
\begin{center}
\mbox{
\includegraphics[width=8cm]{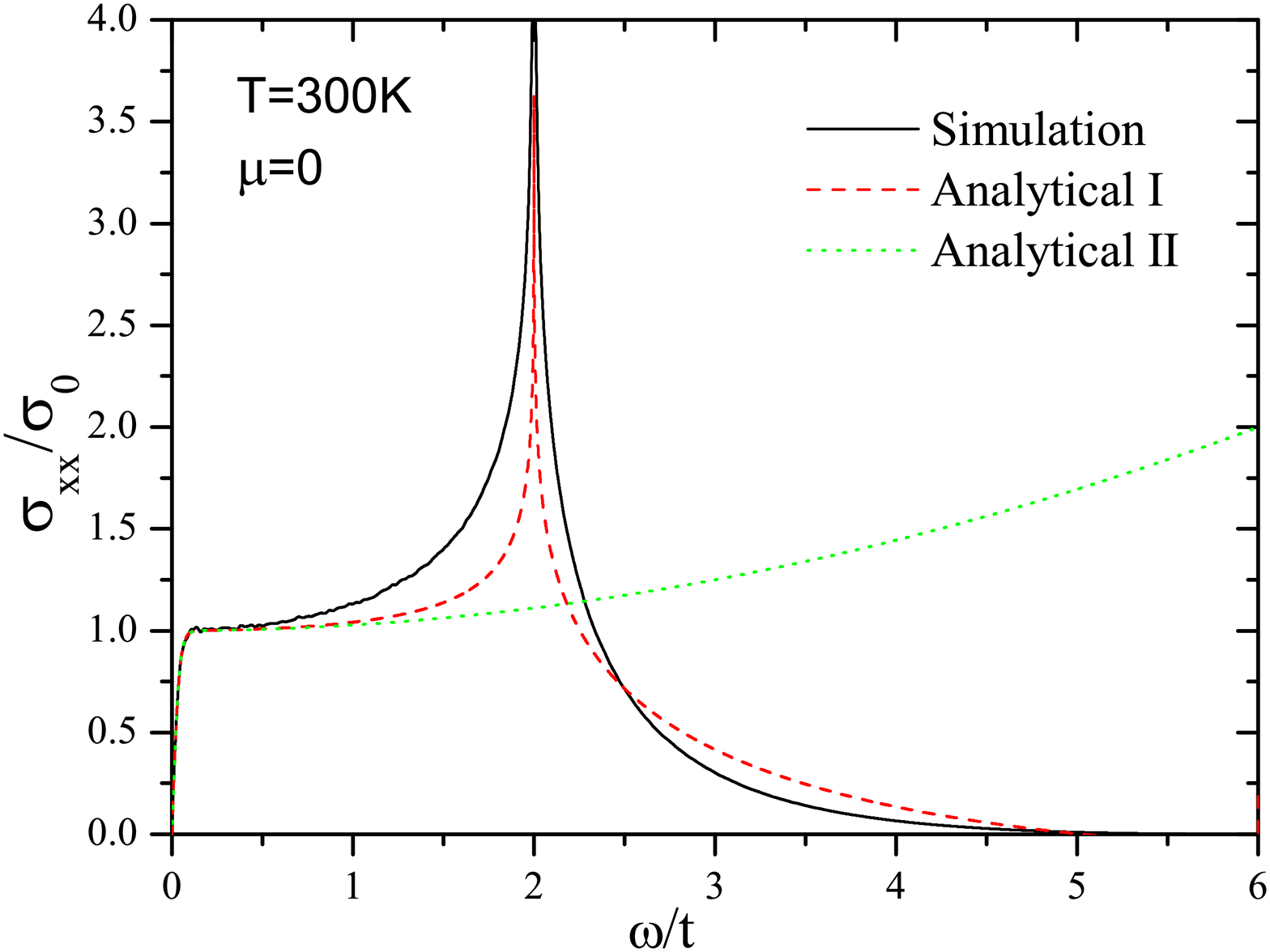}
\includegraphics[width=8cm]{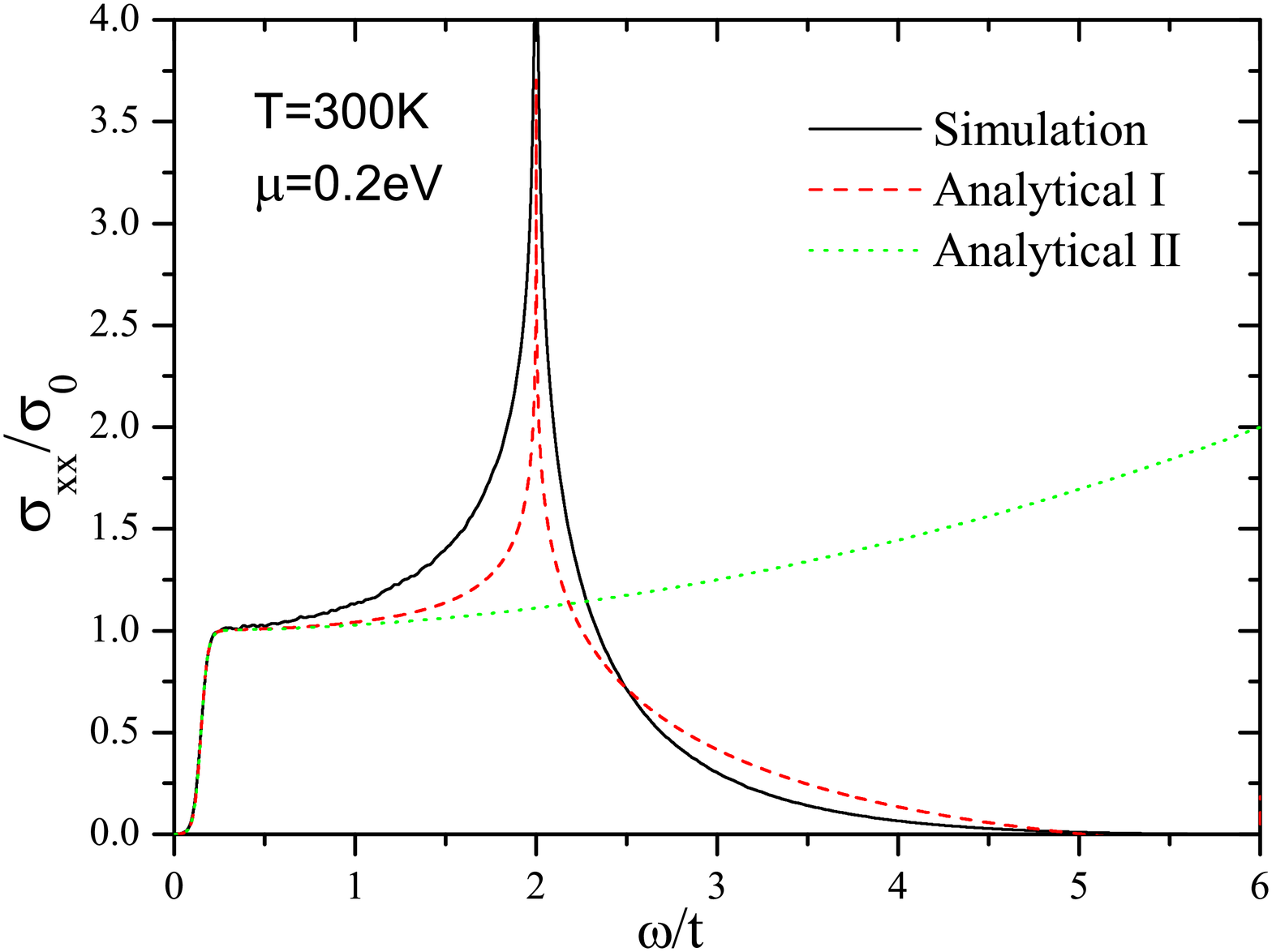}
} 
\mbox{
\includegraphics[width=8cm]{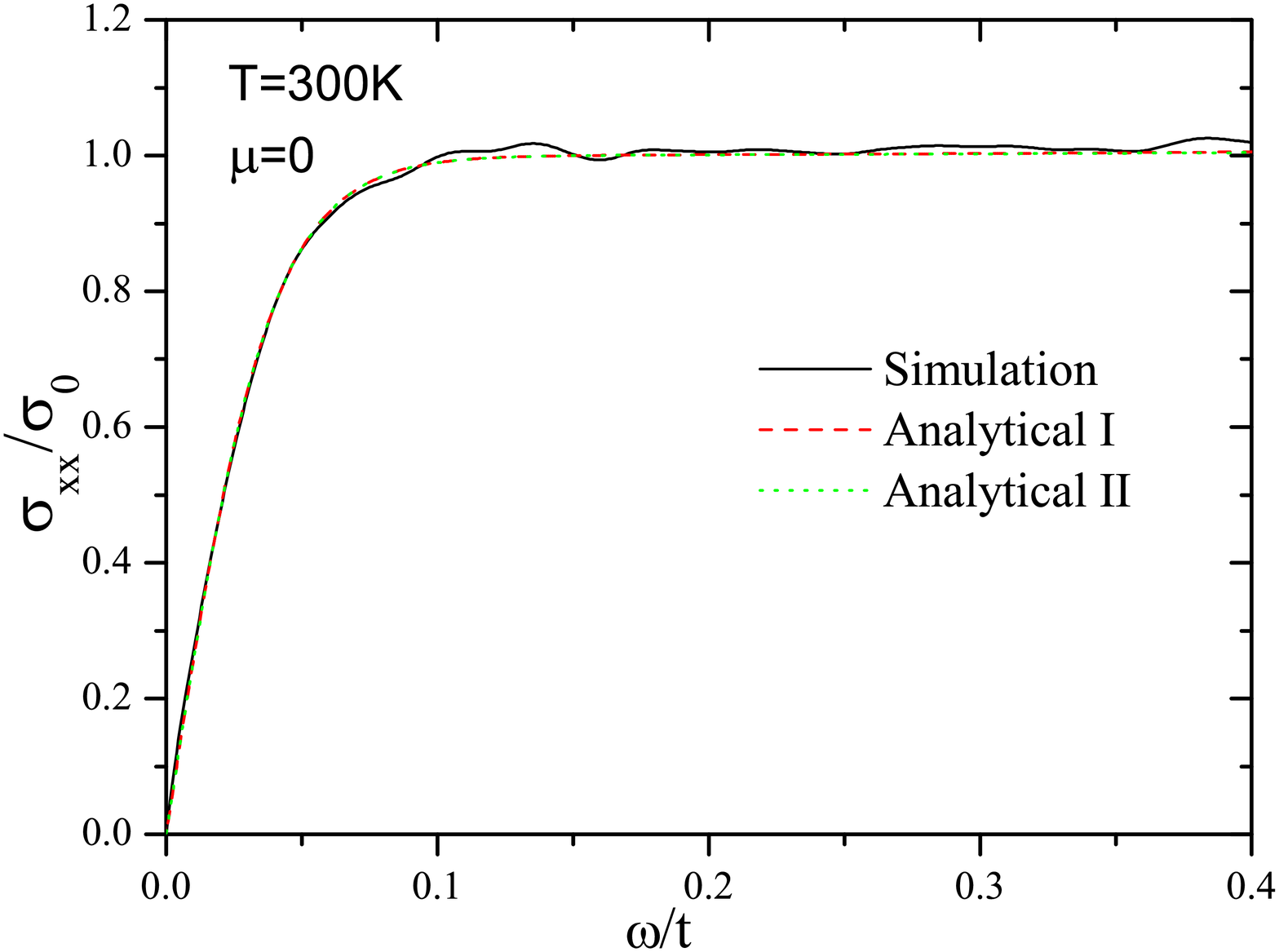}
\includegraphics[width=8cm]{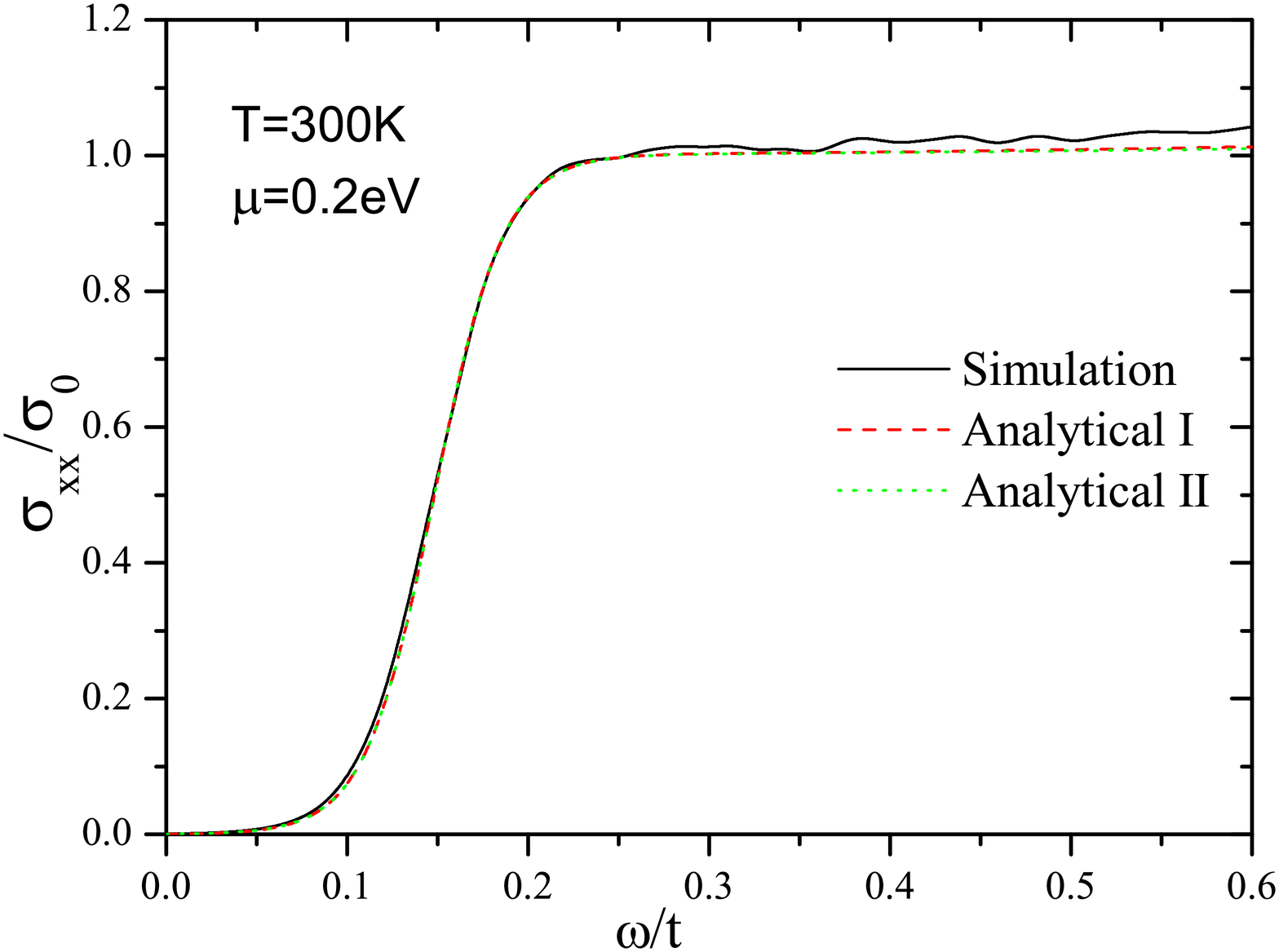}
}
\end{center}
\caption{(Color online) Comparison of the numerically calculated optical
conductivity ($\protect\mu =0$ or $0.2$eV$,$ $T=300K$) with Eq.~(\protect\ref%
{optical_stauber}) (analytical I) and Eq.~(\protect\ref{optical_falkovsky})
(analytical II). The size of the system is $M=N=8192$.}
\label{accompare}
\end{figure*}

\begin{figure}[t]
\begin{center}
\mbox{
\includegraphics[width=8cm]{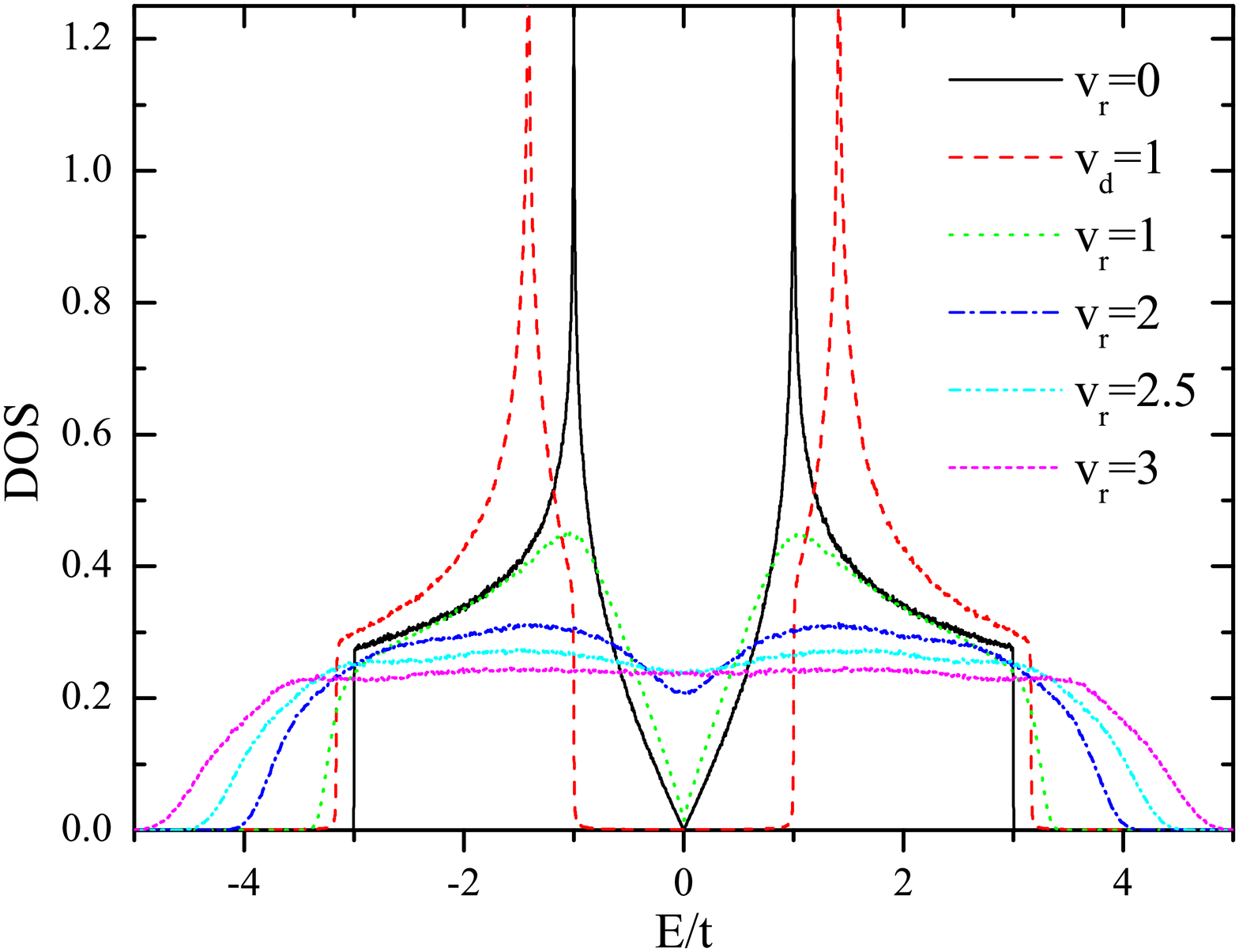}
} \mbox{
\includegraphics[width=8cm]{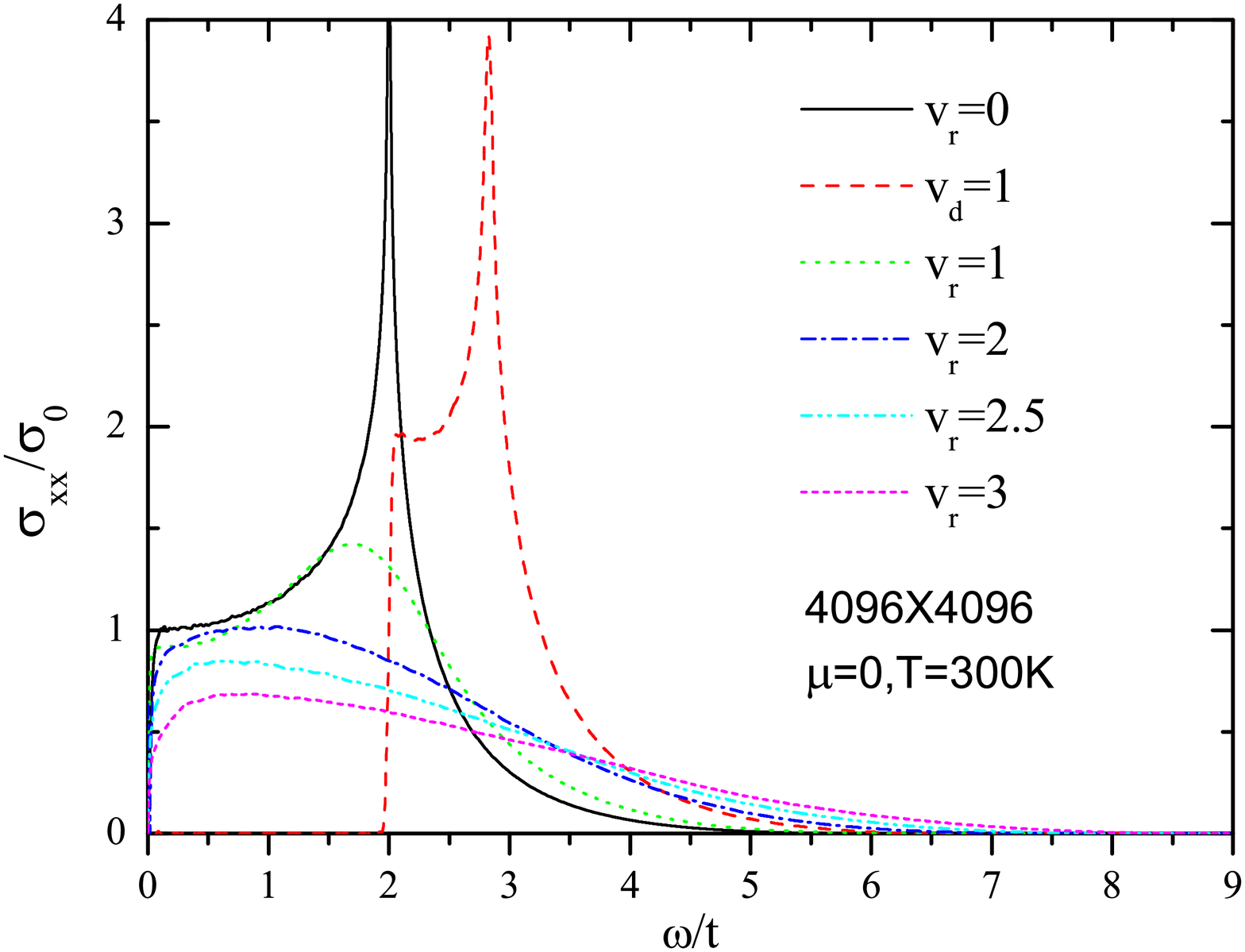}
}
\end{center}
\caption{(Color online) Comparison of DOS (in units of $1/t$) and optical conductivity ($\protect%
\mu =0,T=300K$) with symmetrical random ($v_{r}$) or antisymmetrical fixed ($%
\pm v_{d}$) potential on sublattices A and B. The size of the system is $%
M=N=4096$.}
\label{aca}
\end{figure}

\begin{figure}[t]
\begin{center}
\includegraphics[width=8cm]{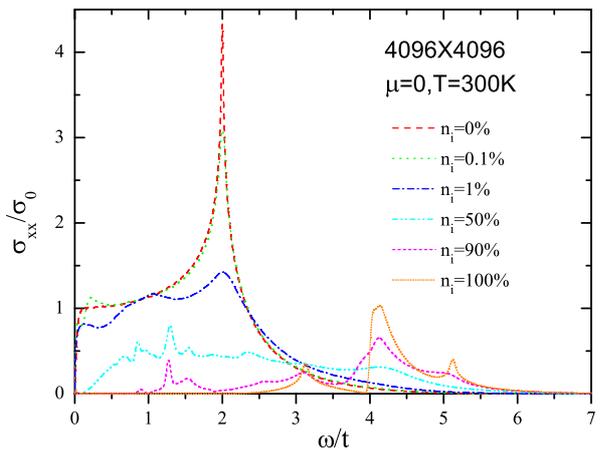}
\end{center}
\caption{(Color online) Comparison of optical conductivity ($\protect\mu %
=0,T=300K$) with different concentration of hydrogen impurities. The size of
the system is $M=N=4096$, except for the clean graphene ($M=N=8192$). }
\label{acxi}
\end{figure}

Kubo's formula for the optical conductivity can be expressed as~\cite%
{Kubo1957}%
\begin{eqnarray}
\mathbf{\sigma }_{\alpha \beta }\left( \omega \right) &=&\lim_{\varepsilon
\rightarrow 0^{+}}\frac{1}{\left( \omega +i\varepsilon \right) \Omega }%
{\Huge \{}-i\left\langle \left[ P_{\alpha },J_{\beta }\right] \right\rangle 
\notag  \label{optical1} \\
&&+\int_{0}^{\infty }e^{i\left( \omega +i\varepsilon \right) t}dt\text{ }%
\left\langle \left[ J_{\alpha }\left( t\right) ,J_{\beta }\right]
\right\rangle {\Huge \}},
\end{eqnarray}%
where $P$ is the polarization operator 
\begin{equation}
P=e\sum_{i}\mathbf{r}_{i}c_{i}^{+}c_{i},
\end{equation}%
and $J$ is the current operator%
\begin{equation}
J=\overset{\cdot }{P}=e\sum_{i}\overset{\cdot }{\mathbf{r}}%
_{i}c_{i}^{+}c_{i}=\frac{i}{\hbar }\left[ H,P\right] .
\end{equation}%
For a generic tight binding Hamiltonian, the current operator can be written
as%
\begin{equation}
J=-\frac{ie}{\hbar }\sum_{i,j}t_{ij}\left( \mathbf{r}_{j}-\mathbf{r}%
_{i}\right) c_{i}^{+}c_{j},
\end{equation}%
and 
\begin{equation}
\left[ P_{\alpha }\mathbf{,}J_{\beta }\right] =-\frac{ie^{2}}{\hbar }%
\sum_{i,j}t_{ij}\left[ \left( \mathbf{r}_{i}-\mathbf{r}_{j}\right) _{\alpha
}\left( \mathbf{r}_{j}-\mathbf{r}_{i}\right) _{\beta }\right] c_{i}^{+}c_{j}.
\end{equation}

The ensemble average in Eq.~(\ref{optical1}) is over the Gibbs distribution,
and the electric field is given by $\mathbf{E}\left( t\right) =\mathbf{E}%
_{0}\exp \left( i\omega +\varepsilon \right) t$ ($\varepsilon $ is a small
parameter introduced in order that $\mathbf{E}\left( t\right) \rightarrow 0$
for $t\rightarrow -\infty $). In graphene, $P$ and $J$ are two-dimensional
vectors, and $\Omega $ is replaced by the area of the sample $S$.

In general, the real part of the optical conductivity contains two parts,
the Drude weight $D$ ($\omega =0$) and the regular part ($\omega \neq 0$).
We omit the calculation of the Drude weight, and focus on the regular part.
For non-interacting electrons, the regular part is \cite{Isihara1971} 
\begin{eqnarray}
\text{Re}\sigma _{\alpha \beta }\left( \omega \right) &=&\lim_{\varepsilon
\rightarrow 0^{+}}\frac{e^{-\beta \hbar \omega }-1}{\hbar \omega \Omega }%
\int_{0}^{\infty }e^{-\varepsilon t}\sin \omega t  \notag  \label{gabw2} \\
&&\times 2\text{Im}\left\langle f\left( H\right) J_{\alpha }\left( t\right) %
\left[ 1-f\left( H\right) \right] J_{\beta }\right\rangle dt,  \notag \\
&&
\end{eqnarray}%
where $\beta =1/k_{B}T,$ $\mu $ is the chemical potential, and the
Fermi-Dirac distribution operator 
\begin{equation}
f\left( H\right) =\frac{1}{e^{\beta \left( H-\mu \right) }+1}.
\end{equation}%

In the numerical calculations, the average in Eq.~(\ref{gabw2}) is performed
over a random phase superposition of all the basis states in the real space,
i.e., the same initial state $\left\vert \varphi \left( 0\right)
\right\rangle $ in calculation of DOS. The Fermi distribution operator $%
f\left( H\right) $ and $1-f\left( H\right) $ can be obtained by the standard
Chebyshev polynomial decomposition (see Appendix B).

By introducing the three wave functions \cite{Iitaka1997} 
\begin{eqnarray}
\left\vert \varphi _{1}\left( t\right) \right\rangle _{x} &=&e^{-\frac{iHt}{%
\hbar }}\left[ 1-f\left( H\right) \right] J_{x}\left\vert \varphi
\right\rangle , \\
\left\vert \varphi _{1}\left( t\right) \right\rangle _{y} &=&e^{-\frac{iHt}{%
\hbar }}\left[ 1-f\left( H\right) \right] J_{y}\left\vert \varphi
\right\rangle , \\
\left\vert \varphi _{2}\left( t\right) \right\rangle &=&e^{-\frac{iHt}{\hbar 
}}f\left( H\right) \left\vert \varphi \right\rangle ,
\end{eqnarray}%
we get all elements of the regular part of Re$\sigma _{\alpha \beta }\left(
\omega \right) $: 
\begin{eqnarray}
\text{Re}\sigma _{\alpha \beta }\left( \omega \right) &=&\lim_{\varepsilon
\rightarrow 0^{+}}\frac{e^{-\beta \hbar \omega }-1}{\hbar \omega \Omega }%
\int_{0}^{\infty }e^{-\varepsilon t}\sin \omega t  \notag \\
&&\times \left[ 2\text{Im}\text{ }\left\langle \varphi _{2}\left( t\right)
\left\vert J_{\alpha }\right\vert \varphi _{1}\left( t\right) \right\rangle
_{\beta }\right] dt.
\end{eqnarray}

\subsection{Optical Conductivity of Clean Graphene}

In Fig.~\ref{accompare}, we compare our numerical results to the analytical
results obtained in Refs.~\onlinecite{Stauber2008,Falkovsky2007,Kuzmenko2008}%
, where the real part of the conductivity in the visible region has the form 
\cite{Stauber2008}%
\begin{eqnarray}
\text{Re}\sigma _{xx} &=&\sigma _{0}\left[ \frac{\pi t^{2}a^{2}}{8A_{c}\hbar
\omega }\rho \left( \frac{\hbar \omega }{2}\right) \left( 18-\frac{\hbar
^{2}\omega ^{2}}{t^{2}}\right) +\frac{\hbar ^{2}\omega ^{2}}{4!2^{4}t^{2}}%
\right]  \notag  \label{optical_stauber} \\
&&\left( \tanh \frac{\hbar \omega +2\mu }{4k_{B}T}+\tanh \frac{\hbar \omega
-2\mu }{4k_{B}T}\right) ,  \notag \\
&&
\end{eqnarray}%
with the minimum conductivity $\sigma _{0}=\pi e^{2}/2h$. Around $\omega =0$
the real part of the conductivity can be simplified as \cite%
{Stauber2008,Falkovsky2007,Kuzmenko2008}%
\begin{eqnarray}
\text{Re}\sigma _{xx} &=&\sigma _{0}\left( \frac{1}{2}+\frac{1}{72}\frac{%
\hbar ^{2}\omega ^{2}}{t^{2}}\right)  \notag  \label{optical_falkovsky} \\
&&\left( \tanh \frac{\hbar \omega +2\mu }{4k_{B}T}+\tanh \frac{\hbar \omega
-2\mu }{4k_{B}T}\right) .
\end{eqnarray}

As we can see from Fig.~\ref{accompare}, the numerical and analytical
results match very well in the low frequency region, but not in the high
frequence region. This is because the analytical expressions are partially
based on the Dirac-cone approximation, i.e., the graphene energy bands are
linearly dependent on the amplitude of the wave vector. It is exact for the
calculations of the low-frequence optical conductivity, but not for
high-frequence. Our numerical method does not use such approximation and has
the same accuracy in the whole spectrum. Furthermore, our numerical results
also show that the conductivity of Re$\sigma _{xx}$ with $\mu =0$ in the
limit of $\omega =0$\ converges to the minimum conductivity $\sigma _{0}$
when the temperature $T\rightarrow 0$.

\subsection{Optical Conductivity of Graphene with Random on-Site Potentials}

The on-site potential disorder can change the electronic properties of
graphene dramatically. For example, if the potentials on sublattices A and B
are not symmetric, a band gap will appear. If we set $v_{d}$ and $-v_{d}$ as
the on-site potential on sublattice A and B, respectively, then a band gap
of size $2v_{d}$ is observed in the central part of DOS and the optical
conductivity in the region $0<\omega <2v_{d}$ becomes zero, see the red
dashed lines ($v_{d}=t$) in Fig.~(\ref{aca}). If the potentials on
sublattice A and B are both uniformly random in a range $[-v_{r},v_{r}]$,
then the spectrum is broaden symmetrically around the neutrality point
(because of the random character of the potentials on sublattice A and B),
and there is no band gap, see the colored lines (except the red one) in
Fig.~(\ref{aca}). It softens the singularities in the DOS, the smearing
being larger for a larger degree of disorder. The smearing of the DOS leads
to the smearing of the optical conductivity, see $\sigma _{xx}$ in Fig.~\ref%
{aca}.

\subsection{Optical Conductivity of Graphene with Resonant Impurities}

In Fig.~\ref{acxi} we present the optical conductivity of graphene with
various concentrations of hydrogen impurities. Small concentrations of the
impurities have a small effect on the optical conductivity, but higher
concentrations change the optical properties dramatically, especially when
the concentration reaches the maximum ($100\%$), i.e., when graphene becomes
graphane. Graphane has a band gap ($2t$), see bottom panel in Fig.~\ref%
{doslargximp}, which leads to the zero optical conductivities within the
region $\left\vert \omega \right\vert \in \lbrack 0,2t]$, see Fig.~\ref{acxi}%
) for $n_{i}=100\%$. At intermediate concentrations, one can clearly see
additional features in the optical conductivity related with the formation
of impurity band. The Van Hove singularity of clean graphene is smeared out
completely for concentrations as small as 1\%.

\section{DC Conductivity}

\begin{figure*}[t]
\begin{center}
\includegraphics[width=16cm]{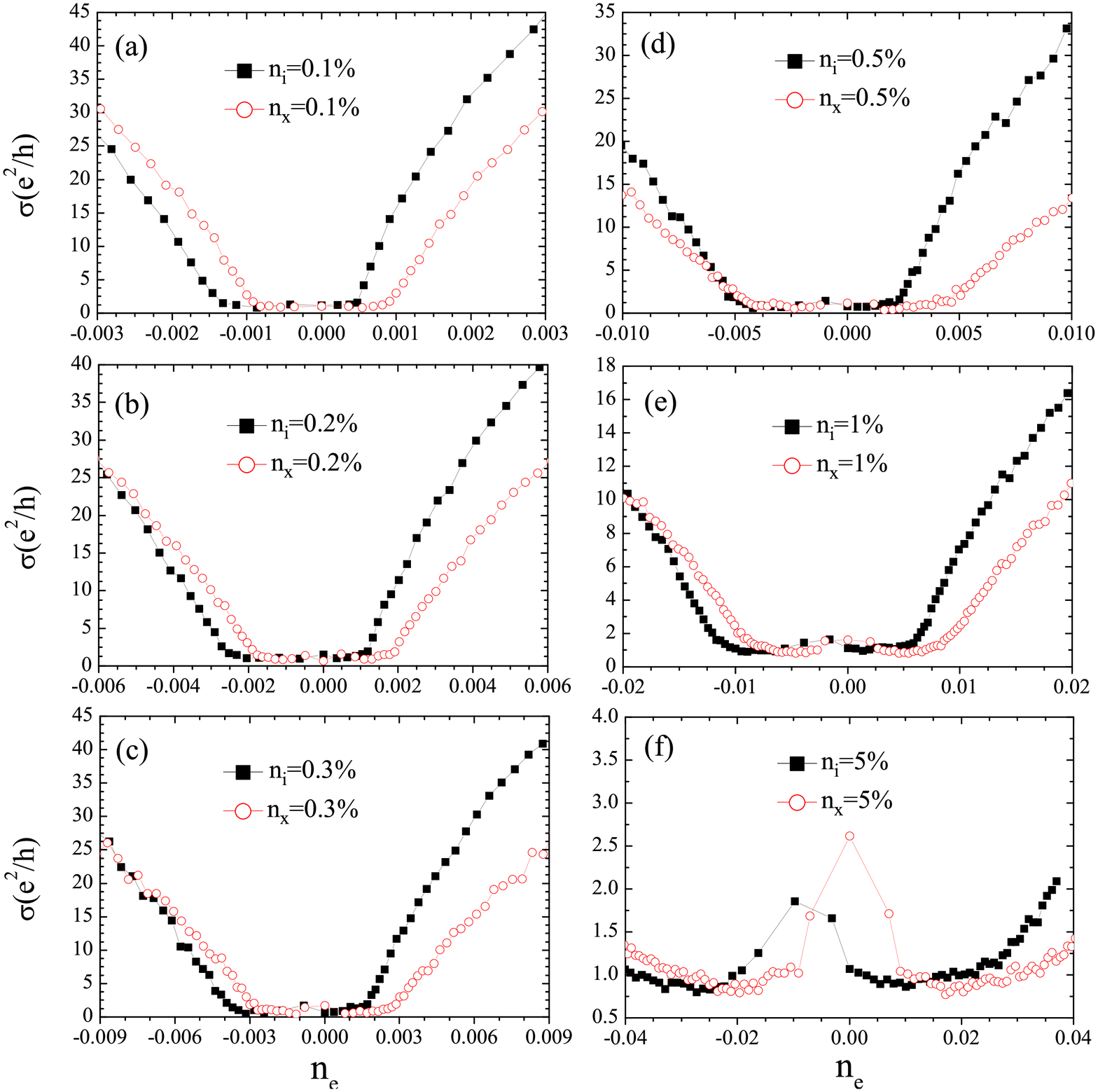}
\end{center}
\caption{(Color online) Conductivity $\protect\sigma $ (in units of $e^{2}/h$%
) as a function of charge carrier concentration $n_{e}$ (in units of
electrons per atom) for different resonant impurity ($\protect\varepsilon %
_{d}=-t/16,$ $V=2t$) or vacancy concentrations ($n_{x}$) : (a) $%
n_{i}=n_{x}=0.1\%,$ (b) $0.2\%,$ (c) $0.3\%,$ (d) $0.5\%$, (e) $1\%,$ (f) $%
5\%$. Numerical calculations are performed on samples containing (a) $%
8192\times 8192$ and (b-f) $4096\times 4096$ carbon atoms. The charge
carrier concentrations $n_{e}$ are obtained by the integral of the
corresponding density of states represented in Fig. \protect\ref{dosximpandx}%
.}
\label{xandximpall}
\end{figure*}

\begin{figure}[t]
\begin{center}
\mbox{
\includegraphics[width=8cm]{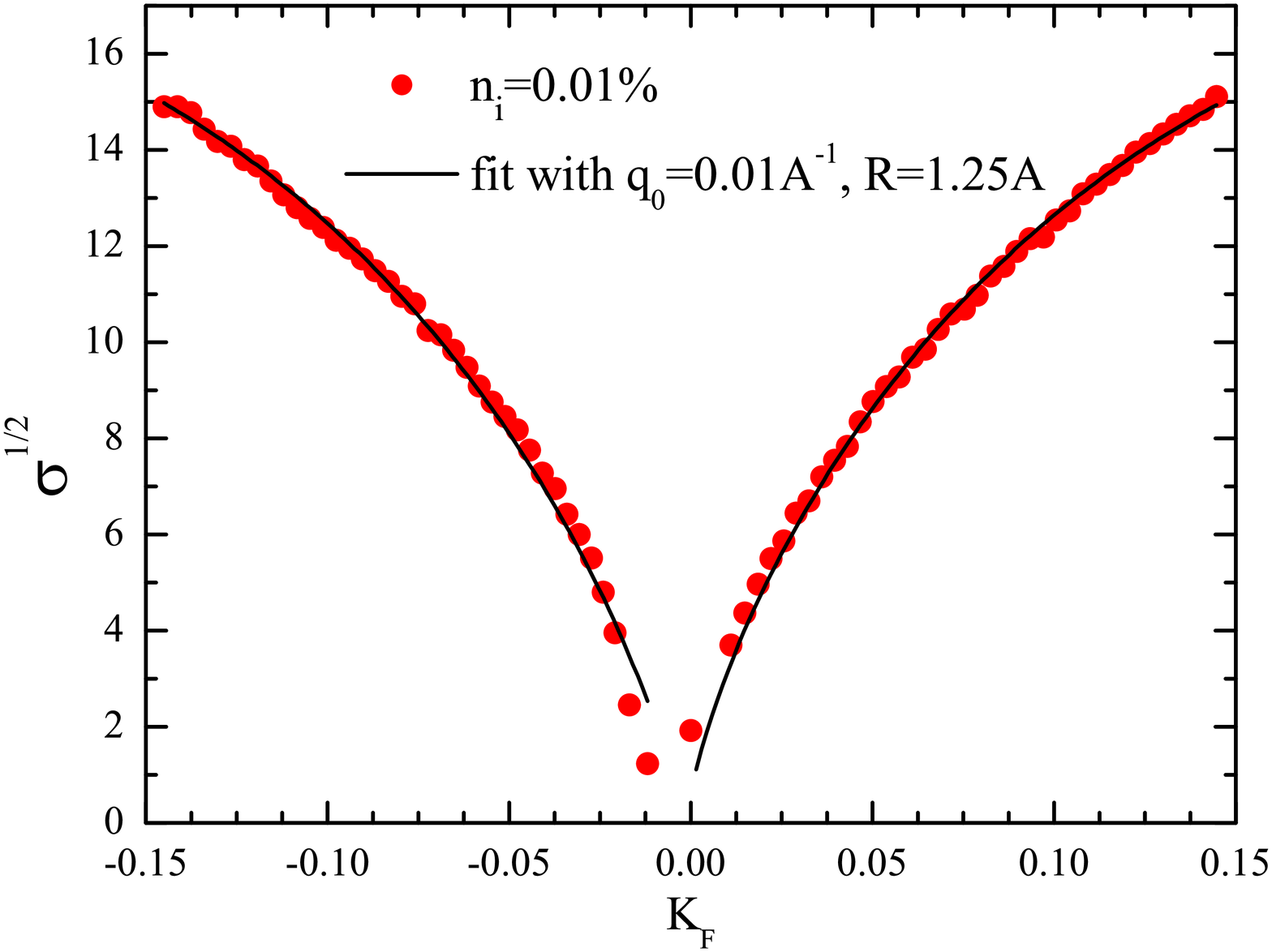}
} \mbox{
\includegraphics[width=8cm]{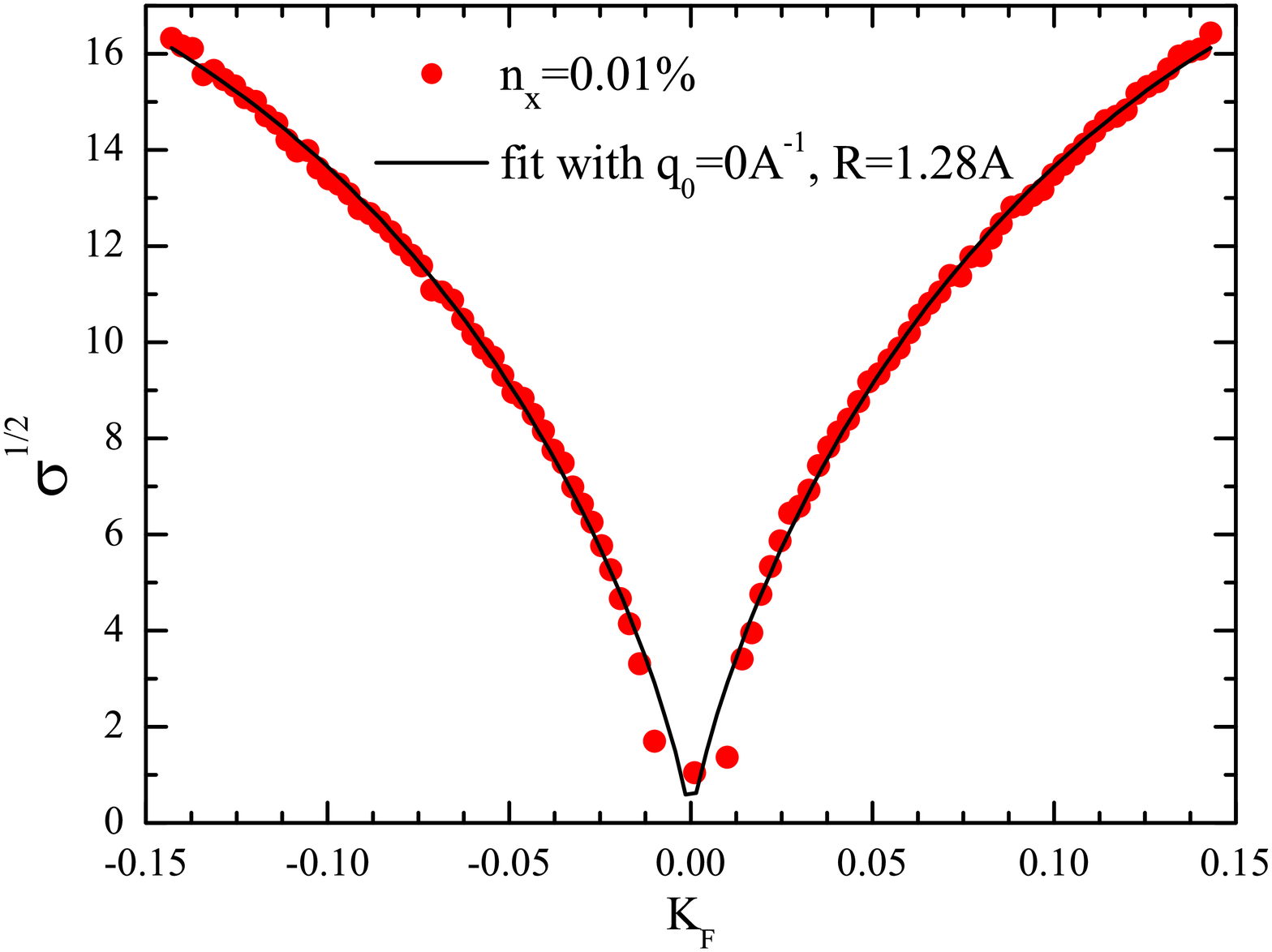}
}
\end{center}
\caption{(Color online) Red dots: conductivity $\protect\sigma $ (in units
of $e^{2}/h$) as a function of $K_{F}$ (in units of \AA $^{-1}$) for
resonant impurity (top panel, $\protect\varepsilon _{d}=-t/16,$ $V=2t$) or
vacancy (bottom panel). The concentration of the impurities is $%
n_{i}=n_{x}=0.01\%.$ Numerical calculations are performed on samples
containing $4096\times 4096$ carbon atoms. Black lines: fit of Eq. (\protect
\ref{fitting}) with $q_{0}=0.01\text{\AA }^{-1}$, $R=1.25\text{\AA }$ for $%
n_{i}=0.01\%$, and $q_{0}=0$, $R=1.28\text{\AA }$ for $n_{x}=0.01\%$.}
\label{sigmafit}
\end{figure}

The DC conductivity can be obtained by taking $\omega \rightarrow 0$ in Eq.~(%
\ref{optical1}) yielding~\cite{Isihara1971} 
\begin{equation}
\mathbf{\sigma }=-\frac{1}{V}Tr\left\{ \frac{\partial f}{\partial H}%
\int_{0}^{\infty }dt\frac{1}{2}\left[ JJ\left( t\right) +J\left( t\right) J%
\right] \right\} .  \label{conduc2}
\end{equation}%
We can use the same algorithm as we used for the optical conductivity to
perform the integration in Eq.~(\ref{conduc2}), but it is not the best
practical way since it only leads to the DC conductivity with one chemical
potential each time, and the number of non-zero terms in Chebyshev
polynomial representation growth exponentially when the temperature tends to
zero. In fact, at zero temperature $\frac{\partial f}{\partial H}$ can be
simplified as 
\begin{equation}
-\frac{\partial f}{\partial H}=\delta \left( E_{F}-H\right) ,
\end{equation}%
and therefore Eq.~(\ref{optical1}) can be simplified as%
\begin{eqnarray}
\mathbf{\sigma }_{T=0} &=&\frac{\pi }{NV}\text{Re}\sum_{m,n=1}^{N}\left%
\langle n\right\vert J\left\vert m\right\rangle \left\langle m\right\vert
J\left\vert n\right\rangle  \notag  \label{conduct0} \\
&&\times \delta \left( E_{F}-E_{m}\right) \delta \left( E_{F}-E_{n}\right) .
\end{eqnarray}%
By using the quasieigenstates $\left\vert \Psi \left( \varepsilon \right)
\right\rangle $ obtained from the spectrum method in Eq.~(\ref{Eigenstate1}%
), we can prove that (see Appendix C) 
\begin{eqnarray}
\mathbf{\sigma } &=&\frac{\rho \left( \varepsilon \right) }{V}%
\int_{0}^{\infty }dt\text{Re}\left[ e^{-i\varepsilon t}\left\langle \varphi
\right\vert Je^{iHt}J\left\vert \varepsilon \right\rangle \right] \approx 
\mathbf{\sigma }_{T=0},  \notag  \label{conducappr} \\
&&
\end{eqnarray}%
where $\left\vert \varphi \right\rangle $ is the same initial random
superposition state as in Eq.~(\ref{phit}) and%
\begin{equation}
\left\vert \varepsilon \right\rangle =\frac{1}{\left\vert \left\langle
\varphi |\Psi \left( \varepsilon \right) \right\rangle \right\vert }%
\left\vert \Psi \left( \varepsilon \right) \right\rangle .
\label{quasieigenstate}
\end{equation}

The accuracy of the quasieigenstates in Eq.~(\ref{Eigenstate1}) are mainly
determined by the time interval and total time steps used in the Fourier
transform. The main limitation of the numerical calculations using Eq.~(\ref%
{Eigenstate1}) is the size of the physical memory that can be used to store
the quasieigenstates $\left\vert \Psi \left( \varepsilon \right)
\right\rangle $. %\subsection{DC Conductivity of Graphene with Impurities}

We used the algorithm presented above to calculate the DC conductivity of
single layer graphene with vacancies or resonant impurities. The results are
shown in Fig.~\ref{xandximpall}. As we can see from the numerical results,
there is plateau of the order of the minimum conductivity \cite%
{Katsnelson2006} $4e^{2}/\pi h$ in the vicinity of the neutrality point, in
agreement with theoretical expectations \cite{Ostrovsky2010}. Finite
concentrations of resonant impurities lead to the formation of a low energy
impurity band (see increased DOS at low energies in Fig. \ref{dosximpandx}).
At impurity concentrations of the order of a few percent (Fig. \ref%
{xandximpall} e, f) this impurity band contributes to the conductivity and
can lead to a maximum of $\sigma $ in the midgap region. The impurity band
can host two electrons per impurity. For impurity concentrations below $\sim
5\%$, this leads to a plateau-shaped minimum of width $2n_{i}$ (or $2n_{x}$)
in the conductivity vs. $n_{e}$ curves around the neutrality point.
Analyzing experimental data of the plateau width (similar to the analysis
for N$_{2}$O$_{4}$ acceptor states in Ref. \onlinecite{Wehling2008}) can
therefore yield an independent estimate of impurity concentration.

Beyond the plateau around the neutrality point, the conductivity is
inversely proportional to the concentration of the impurities, and
approximately proportional to the carrier concentration $n_{e}$. This is
consistent with the approach based on the Boltzmann equation, which in the
limit of resonant impurities with $V\rightarrow \infty $, yields for the
conductivity~\cite{Katsnelson2007,Katsnelson2008,Robinson2008,Wehling2010} 
\begin{equation}
\sigma \approx (2e^{2}/h)\frac{2}{\pi }\frac{n_{e}}{n_{i}}\ln ^{2}\left\vert 
\frac{E}{D}\right\vert ,  \label{eqn:imp_lim}
\end{equation}%
where $n_{e}=E_{F}^{2}/D^{2}$ is the number of charge carriers per carbon
atom, and $D$ is of order of the bandwidth. Equantion (\ref{eqn:imp_lim})
yields the same behavior as for vacancies \cite{Stauber2007}. Note that for
the case of the resonance shifted with respect to the neutrality point the
consideration of Ref. \onlinecite{Katsnelson2007} leads to the dependence 
\begin{equation}
\sigma \propto \left( q_{0}\pm k_{F}\ln {k_{F}R}\right) ^{2},
\label{fitting}
\end{equation}%
where $\pm $ corresponds to electron and hole doping, respectively, and $R$
is the effective impurity radius. The Boltzmann approach does not work near
the neutrality point where quantum corrections are dominant \cite%
{Ostrovsky2006, Katsnelson2006, Auslender2007}. In the range of
concentrations, where the Boltzmann approach is applicable the conductivity
as a function of energy fits very well to the dependence given by Eq. (\ref%
{fitting}), as for example shown in Fig. (\ref{sigmafit}), with $q_{0}=0.01%
\text{\AA }^{-1}$, $R=1.25\text{\AA }$ for $n_{i}=0.01\%$, and $q_{0}=0$, $%
R=1.28\text{\AA }$ for $n_{x}=0.01\%$. The relation of these results to
experiment is discussed in Ref. \onlinecite{Wehling2010}.

The advantage of the method used here for the calculation of the DC
conductivity is that the results do not depend on the upper time limit in
the integration since the contributions to the integrand in Eq.(\ref%
{conducappr}) corresponding to different energies tends to zero fast enough
when the time is large. The propagation time for the integration depends on
the concentration of the disorder, i.e., larger concentration leads to
faster decay of the corrections. The disadvantage of this method is that a
lot of memory may be needed to store the coefficients of many
quasieigenstates. Furthermore, since $\left\vert \varepsilon \right\rangle $
in Eq.~(\ref{quasieigenstate}) contains the factor $1/\left\vert
\left\langle \varphi |\Psi \left( \varepsilon \right) \right\rangle
\right\vert $, this may cause problems when $\left\vert \left\langle \varphi
|\Psi \left( \varepsilon \right) \right\rangle \right\vert $ is very small.
For example, when using this method to calculate the Hall conductivity in
the presence of strong magnetic fields, tiny $\left\vert \left\langle
\varphi |\Psi \left( \varepsilon \right) \right\rangle \right\vert $ (out
the Landau levels) will leads to large fluctuations of the calculated
conductivity. Nevertheless, the conductivities without the presence on the
magnetic filed in our paper are agreement with the results reported in Refs. %
\onlinecite{Bang2010} (hydrogenated graphene) and \onlinecite{Wu2010}
(graphene with vacancies), and both papers are based on the numerical
calculation of the Kubo-Greenwood formula, as proposed in Ref.%
\onlinecite{Roche1997}. To calculate the Hall conductivity accurately our
method should be developed further.

\section{Summary}

We have presented a detailed numerical study of the electronic properties of
single-layer graphene with resonant (\textquotedblleft
hydrogen\textquotedblright ) impurities and vacancies within a framework of
noninteracting tight-binding model on the honeycomb lattice. The algorithms
developed in this paper are based on the numerical solution of the
time-dependent Schr\"{o}dinger equation, the fundamental operation being the
action of the evolution operator on a general wave vector. We do not need to
diagonalize the Hamiltonian matrix to obtain the eigenstates and therefore
the method can be applied to very large crystallites which contains millions
of atoms. Furthermore since the operation of the Hamiltonian matrix on a
general wave vector does not require any special symmetry of the matrix
elements, this flexibility can be exploited to study different kinds of
disorder and impurities in the noninteracting tight-binding model.

The algorithms for the calculation of density of states, quasieigenstates,
AC and DC conductivities, are applicable to any 1D, 2D and 3D lattice
structure, not only to a single layer of carbon atoms arranged in a
honeycomb lattice. The calculation for the electronic properties of
multilayer graphene can be easily obtained by adding the hoping between the
corresponding atoms of different layers.

Our computational results give a consistent picture of behavior of the
electronic structure and transport properties of functionalized graphene in
a broad range of concentration of impurities (from graphene to graphane).
Formation of impurity bands is the main factor determining electrical and
optical properties at intermediate impurity concentrations, together with
the appearance of a gap near the graphane limit.

\section{Acknowledgement}

The support by the Stichting Fundamenteel Onderzoek der Materie (FOM) and
the Netherlands National Computing Facilities foundation (NCF) are
acknowledged.

\section{Appendix A}

Suppose $x\in \left[ -1,1\right] $, then 
\begin{equation}
e^{-izx}=J_{0}(z)+2\sum_{m=1}^{\infty }\left( -i\right) ^{m}J_{m}\left(
z\right) T_{m}\left( x\right) ,  \label{exp2}
\end{equation}%
where $J_{m}(z)$ is the Bessel function of integer order $m$, and $%
T_{m}\left( x\right) =\cos \left[ m\arccos \left( x\right) \right] $ is the
Chebyshev polynomial of the first kind. $T_{m}\left( x\right) $ obeys the
following recurrence relation:%
\begin{equation}
T_{m+1}\left( x\right) +T_{m-1}\left( x\right) =2xT_{m}\left( x\right) .
\label{Cheb1}
\end{equation}

Since the Hamiltonian $H$ has a complete set of eigenvectors $\left\vert
E_{n}\right\rangle $ with real valued eigenvalues $E_{n}$, we can expand the
wave function $\left\vert \phi (0)\right\rangle $ as a superposition of the
eigenstates $\left\vert n\right\rangle $ of $H$ 
\begin{equation}
\left\vert \phi (0)\right\rangle =\sum_{n=1}^{N}\left\vert n\right\rangle
\left\langle n|\phi (0)\right\rangle ,
\end{equation}%
and therefore%
\begin{equation}
\left\vert \phi (t)\right\rangle =e^{-itH}\left\vert \phi (0)\right\rangle
=\sum_{n=1}^{N}e^{-itE_{n}}\left\vert n\right\rangle \left\langle n|\phi
(0)\right\rangle .  \label{Phi0}
\end{equation}%
By using the inequality%
\begin{equation}
\left\Vert \sum X_{n}\right\Vert \leq \sum \left\Vert X_{n}\right\Vert ,
\end{equation}%
with the Hamiltonian $H$ of Eq.(\ref{Hamiltonian}) we find%
\begin{eqnarray}
\left\Vert H\right\Vert _{b} &\equiv &3t_{\max }+6t_{\max }^{\prime
}+\left\vert v\right\vert _{\max }+\left\vert \varepsilon _{d}\right\vert
+\left\vert V\right\vert  \notag \\
&\geqslant &\max \{E_{n}\}.
\end{eqnarray}%
Introduce new variables $\hat{t}\equiv t\left\Vert H\right\Vert _{b}$\ and $%
\hat{E}_{n}\equiv E_{n}/\left\Vert H\right\Vert _{b}$, where $\hat{E}_{n}$
are the eigenvalues of a modified Hamiltonian $\hat{H}\equiv H/\left\Vert
H\right\Vert _{b}$, that is%
\begin{equation}
\hat{H}\left\vert E_{n}\right\rangle =\hat{E}_{n}\left\vert
E_{n}\right\rangle .
\end{equation}

By using Eq.~(\ref{exp2}), the time evolution of $\left\vert \phi
(t)\right\rangle $ can be represented as 
\begin{eqnarray}
\left\vert \phi (t)\right\rangle &=&\left[ J_{0}(\hat{t})\hat{T}_{0}\left( 
\hat{H}\right) +2\sum_{m=1}^{\infty }J_{m}\left( \hat{t}\right) \hat{T}%
_{m}\left( \hat{H}\right) \right] \left\vert \phi (0)\right\rangle ,  \notag
\\
&&
\end{eqnarray}%
where the modified Chebyshev polynomial $\hat{T}_{m}\left( \hat{E}%
_{n}\right) $ is 
\begin{equation}
\hat{T}_{m}\left( \hat{E}_{n}\right) =\left( -i\right) ^{m}T_{m}\left( \hat{E%
}_{n}\right) ,
\end{equation}%
obeys the recurrence relation 
\begin{eqnarray}
\hat{T}_{m+1}\left( \hat{H}\right) \left\vert \phi \right\rangle &=&-2i\hat{H%
}\hat{T}_{m}\left( \hat{H}\right) \left\vert \phi \right\rangle +\hat{T}%
_{m-1}\left( \hat{H}\right) \left\vert \phi \right\rangle ,  \notag \\
\hat{T}_{0}\left( \hat{H}\right) \left\vert \phi \right\rangle
&=&I\left\vert \phi \right\rangle ,\hat{T}_{1}\left( \hat{H}\right)
\left\vert \phi \right\rangle =-i\hat{H}\left\vert \phi \right\rangle .
\end{eqnarray}

\section{Appendix B}

In general, a function $f(x)$ whose values are in the range $\left[ -1,1%
\right] $ can be expressed as%
\begin{equation}
f(x)=\frac{1}{2}c_{0}T_{0}\left( x\right) +\sum_{k=1}^{\infty
}c_{k}T_{k}\left( x\right) ,
\end{equation}%
where $T_{k}\left( x\right) =\cos \left( k\arccos x\right) $ and the
coefficients $c_{k}$ are%
\begin{equation}
c_{k}=\frac{2}{\pi }\int_{-1}^{1}\frac{dx}{\sqrt{1-x^{2}}}f\left( x\right)
T_{k}\left( x\right) .
\end{equation}%
Let $x=\cos \theta $, then $T_{k}\left( x\right) =T_{k}\left( \cos \theta
\right) =\cos k\theta $, and%
\begin{eqnarray}
c_{k} &=&\frac{2}{\pi }\int_{0}^{\pi }f\left( \cos \theta \right) \cos
k\theta d\theta  \notag \\
&=&\text{Re}\left[ \frac{2}{N}\sum_{n=0}^{N-1}f\left( \cos \frac{2\pi n}{N}%
\right) e^{\frac{2\pi ink}{N}}\right] ,
\end{eqnarray}%
which can be calculated by the fast Fourier transform.

For the operators $f=ze^{-\beta H}/\left( 1+ze^{-\beta H}\right) $, where $%
z=\exp \left( \beta \mu \right) $ is the fugacity, we normalize $H$ such
that $\widetilde{H}=H\mathbf{/}\left\vert \left\vert H\right\vert
\right\vert $ has eigenvalues in the range $\left[ -1,1\right]$ and put $%
\widetilde{\beta }=\beta \left\vert \left\vert H\right\vert \right\vert$.
Then 
\begin{equation}
f\left( \widetilde{H}\right) =\frac{ze^{-\widetilde{\beta }\widetilde{H}}}{%
1+ze^{-\widetilde{\beta }\widetilde{H}}}=\sum_{k=0}^{\infty
}c_{k}T_{k}\left( \widetilde{H}\right) ,
\end{equation}%
where $c_{k}$ are the Chebyshev expansion coefficients of%
\begin{equation}
f\left( x\right) =\frac{ze^{-\widetilde{\beta }x}}{1+ze^{-\widetilde{\beta }%
x}},
\end{equation}%
and the Chebyshev polynomial $T_{k}\left( \widetilde{H}\right) $ can be
obtained by the recursion relations%
\begin{equation}
T_{k+1}\left( \widetilde{H}\right) -2\widetilde{H}T_{k}\left( \widetilde{H}%
\right) +T_{k-1}\left( \widetilde{H}\right) =0,
\end{equation}%
with%
\begin{equation}
T_{0}\left( \widetilde{H}\right) =1,T_{1}\left( \widetilde{H}\right) =%
\widetilde{H}.
\end{equation}

\section{Appendix C}

The random superposition state (RSS) $\left\vert \varphi \right\rangle $ in
the real space can be represented in the energy eigenbases as%
\begin{equation}
\left\vert \varphi \right\rangle =\sum_{n}A_{n}\left\vert n\right\rangle.
\end{equation}%
By using the expression Eq.~(\ref{Eigenstate1}) of $\left\vert \Psi \left(
\varepsilon \right)\right\rangle $ we obtain 
\begin{equation}
\left\vert \left\langle \varphi |\Psi \left( \varepsilon \right)
\right\rangle \right\vert =\sqrt{\sum_{n}\left\vert A_{n}\right\vert
^{2}\delta \left( E-E_{n}\right) },
\end{equation}%
and 
\begin{equation}
\left\vert \varepsilon \right\rangle =\frac{1}{\sum_{n}\left\vert
A_{n}\right\vert ^{2}\delta \left( \varepsilon -E_{n}\right) }%
\sum_{n}A_{n}\delta \left( \varepsilon -E_{n}\right) \left\vert
n\right\rangle .
\end{equation}%
Therefore the conductivity in Eq.~(\ref{conducappr}) becomes 
\begin{eqnarray}
\mathbf{\sigma } &=&\frac{1}{V}\frac{\rho \left( \varepsilon \right) }{%
\sum_{n}\left\vert A_{n}\right\vert ^{2}\delta \left( \varepsilon
-E_{n}\right) }\int_{0}^{\infty }dt\text{Re}[e^{-i\left( \varepsilon
-E_{m}\right) t}  \notag \\
&&\times \sum_{m,k}A_{k}^{\ast }\left\langle k\right\vert J\left\vert
m\right\rangle \left\langle m\right\vert J\sum_{n}A_{n}\delta \left(
\varepsilon -E_{n}\right) \left\vert n\right\rangle ]  \notag \\
&=&\frac{\pi }{V}\frac{\rho \left( \varepsilon \right) }{\sum_{n}\left\vert
A_{n}\right\vert ^{2}\delta \left( \varepsilon -E_{n}\right) }\text{Re}%
\sum_{m,k,n}A_{n}A_{k}^{\ast }  \notag \\
&&\times \left\langle k\right\vert J\left\vert m\right\rangle \left\langle
m\right\vert J\left\vert n\right\rangle \delta \left( \varepsilon
-E_{m}\right) \delta \left( \varepsilon -E_{n}\right).
\end{eqnarray}%
Dividing $\sum_{m,k,n}$ into two parts with $k=n$ and $k\neq n$, the
conductivity reads 
\begin{eqnarray}
\mathbf{\sigma } &=&\frac{\pi }{V}\frac{\rho \left( \varepsilon \right) }{%
\sum_{n}\left\vert A_{n}\right\vert ^{2}\delta \left( \varepsilon
-E_{n}\right) }\text{Re}\sum_{m,n}\left\vert A_{n}\right\vert ^{2}  \notag \\
&&\times \left\langle n\right\vert J\left\vert m\right\rangle \left\langle
m\right\vert J\left\vert n\right\rangle \delta \left( \varepsilon
-E_{m}\right) \delta \left( \varepsilon -E_{n}\right)  \notag \\
&&+\frac{\pi }{V}\frac{\rho \left( \varepsilon \right) }{\sum_{n}\left\vert
A_{n}\right\vert ^{2}\delta \left( \varepsilon -E_{n}\right) }\text{Re}%
\sum_{m,k\neq n}A_{n}A_{k}^{\ast }  \notag \\
&&\times \left\langle k\right\vert J\left\vert m\right\rangle \left\langle
m\right\vert J\left\vert n\right\rangle \delta \left( \varepsilon
-E_{m}\right) \delta \left( \varepsilon -E_{n}\right) ,  \notag \\
&&
\end{eqnarray}

When the sample size $N\rightarrow \infty $, the RSS in real space is
equivalent to a RSS in the energy basis, and we have $\left\vert
A_{n}\right\vert ^{2}\approx 1/N,$ $\rho \left( \varepsilon \right) \approx
\sum_{n}\left\vert A_{n}\right\vert ^{2}\delta \left( \varepsilon
-E_{n}\right)$. Then the second terms in above expression is close to zero
because of the cancellation of the random complex coefficients $%
A_{n}A_{k}^{\ast }$. Thus, we have proven that 
\begin{eqnarray}
\mathbf{\sigma } &=&\frac{\rho \left( \varepsilon \right) }{V}%
\int_{0}^{\infty }dt\text{Re}\left[ e^{-i\varepsilon t}\left\langle \varphi
\right\vert Je^{iHt}J\left\vert \varepsilon \right\rangle \right]  \notag \\
&\approx &\frac{\pi }{NV}\text{Re}\sum_{m,n}\left\langle n\right\vert
J\left\vert m\right\rangle \left\langle m\right\vert J\left\vert
n\right\rangle \delta \left( \varepsilon -E_{m}\right) \delta \left(
\varepsilon -E_{n}\right) ,  \notag \\
&&
\end{eqnarray}%
which is just Eq.~(\ref{conduct0}).

Introducing%
\begin{equation}
\left\vert \varphi _{1}\left( t\right) \right\rangle
_{x}=e^{-iHt}J_{x}\left\vert \varphi \right\rangle ,\text{ }\left\vert
\varphi _{1}\left( t\right) \right\rangle _{y}=e^{-iHt}J\left\vert \varphi
\right\rangle ,
\end{equation}%
the DC conductivities at zero temperature is given by%
\begin{eqnarray}
\mathbf{\sigma }_{\alpha \beta }\left( \varepsilon ,T=0\right) &=&\frac{1}{V}%
\int_{0}^{\infty }\text{ }\text{Re}\left[ e^{-i\varepsilon t}\text{ }%
_{\alpha }\left\langle \varphi _{1}\left( t\right) \left\vert J_{\beta
}\right\vert \varepsilon \right\rangle \right] dt.  \notag \\
&&
\end{eqnarray}
The DC conductivity for temperature $T>0$ is%
\begin{equation}
\sigma _{\alpha \beta }=\sum_{\varepsilon }\beta \left[ 1-f\left(
\varepsilon \right) \right] f\left( \varepsilon \right) \mathbf{\sigma }%
_{\alpha \beta }\left( \varepsilon ,T=0\right) .
\end{equation}

\end{document}